Querying Labeled Time Series Data with Scenario Programs

by

Devan Shanker

A thesis submitted in partial satisfaction of the

requirements for the degree of

Master of Science

in

Electrical Engineering and Computer Science

in the

Graduate Division

of the

University of California, Berkeley

Committee in charge:

Professor Sanjit A. Seshia, Chair
Professor Alvin Cheung

Spring 2024

Querying Labeled Time Series Data with Scenario Programs

# Abstract

Querying Labeled Time Series Data with Scenario Programs

by

Devan Shanker

Master of Science in Electrical Engineering and Computer Science

University of California, Berkeley

Professor Sanjit A. Seshia, Chair

In order to ensure that autonomous vehicles (AVs) are safe for on-road deployment, simulation-based testing has become an integral complement to on-road testing. The rise in simulation testing and validation reflects a growing need to verify that AV behavior is consistent with desired outcomes even in edge case scenarios — which may seldom or never appear in on-road testing data. This raises a critical question: to what extent are AV failures in simulation consistent with data collected from real-world testing? As a result of the gap between simulated and real sensor data (sim-to-real gap), failures in simulation can either be spurious (simulation- or simulator-specific issues) or relevant (safety-critical AV system issues). One possible method for validating if simulated time series failures are consistent with real-world time series sensor data could involve retrieving instances of the failure scenario from a real-world time series dataset, in order to understand AV performance in these scenarios. Adopting this strategy, we propose a formal definition of what constitutes a match between a real-world labeled time series data item and a simulated scenario written from a fragment of the SCENIC probabilistic programming language for simulation generation. With this definition of a match, we develop a querying algorithm that identifies the subset of a labeled time series dataset matching a given scenario. To allow this approach to be used to verify the safety of other cyber-physical systems (CPS), we present a definition and algorithm for matching in no way limited to the autonomous vehicles domain. Experiments demonstrate the precision and scalability of the algorithm for a set of challenging and uncommon time series scenarios identified from the NUSCENES autonomous driving dataset. We include a full system implementation of the querying algorithm freely available for use across a wide range of CPS.

To S.D., R.S., S.S., N.S.

Whether my work takes the form of formal verification, autonomous vehicles, or multimodal retrieval and generation, to me creating safe and verified AI systems is simply about giving us all a few more moments with the people we love.

# Contents

# List of Figures

vi# List of Tables

2.1 Scalability Test Results for (N) DoUntil Statements (in sec) . . . . . . . . . . . . 31
2.2 Scalability Test Results for (N) Do Statements (in sec) . . . . . . . . . . . . . . . 31
2.3 Scalability Test Results for (N)ested Try-Interrupt (in sec) . . . . . . . . . . . . 31
2.4 Scalability Test Results for Try-(N)terrupt (in sec) . . . . . . . . . . . . . . . . . 31

vii# Acknowledgments

This work would not be possible without the inspiring dedication and mentorship of Eddie Kim and the incredible vision and guidance of Professor Sanjit A. Seshia. It has been a pleasure and an honor to work together towards such a meaningful goal. I am proud to say there is truly nowhere I would have rather spent the past year. I greatly appreciate the advising and engagement from Professor Alvin Cheung during the revision and preparation of this thesis.

I hope to extend a heartfelt thanks to the following individuals who made this thesis possible. Throughout this thesis, I will use the term *we* to honor and recognize the invaluable contributions of the individuals below:

Dr. Eddie Kim, whom I cannot thank enough for incredible work translating SCENIC predicates into SMT formulas, invaluable insights from the precursor to this work, and assistance motivating, formalizing, and designing an algorithm for such a monumental problem. Professor Daniel Fremont and Professor Hazem Torfah, for the assistance designing and validating the algorithm, formalizing the problem statement, and understanding the intricacies of SCENIC and formal verification. Adwait Godbole and Federico Mora, for unmatched UCLID5 prowess and assistance constructing a hierarchical SCENIC program representation. Oliver Ye, Jillian Goldberg, and Adi Bose for adaptability, spirit, and perserverance through the challenges of implementing trace labeling for such an intricate system. Oliver, your contributions to the SCENIC to SMT translation process for predicate abstraction and evaluation was instrumental to the algorithm implementation. Hongbeen Park, for building a creative computer vision trace labeling system with impressive performance.

For each of the concurrent works not included in this thesis (*ScenarioNL*, *RAGA*, *IMAGINE*, *CalChat*), I would like to thank each of my coauthors for their vital contributions. Josh, the passion and dedication to your work is unmatched. It was a pleasure to work with you and build out your visions for *CalChat* and *RAGA*. Tarun, thank you for helping me turn the *IMAGINE* dream into a reality. In the *ScenarioNL* work mentioned in the Introduction (Section 1.1), I would like to thank each of my coauthors for their outstanding work towards the final system. Everything we did would not have been possible without any of you.

A.C., T.A., R.P., A.J., D.Y., you were the best friends I could have ever made through this process. This thesis and each additional work mentioned in this section would not be what they are without your unwavering friendship and support.

Above all, thank you Rohin for being the best housemate, brother, friend, and personal stylist I could ever ask for. Mom and Dad, thank you for teaching me to forge my own path, to never compromise on what is right, and to chase my dreams wherever they take me.

# Chapter 1

# Motivation

## 1.1 Introduction

Simulation-based testing has become a critical component of safety and performance validation of cyber-physical systems (CPS) across domains including indoor robotics [1], unmanned aerial vehicle systems [2], and autonomous vehicles [3], [4]. Considering the example of the autonomous vehicles domain, testing in simulation allows for the reconstruction of high-risk scenarios in a safe, efficient, and scalable manner [5], [6]. Several national initiatives, including the U.S. National Highway Traffic Safety Administration (NHTSA) and the United Nations Inland Transport Committee (UNECE WP.29), have called for rigorous simulation testing of self-driving systems prior to on-road deployment [7]–[10]. Many simulator platforms can render specific types of autonomous vehicle sensor data for testing and validation of specific autonomy components [11]–[13]. Multiple open-source simulation environments support the full automation of the AV testing process and are actively in use for the development of on-road autonomous vehicle systems [14]–[16]. Several techniques recently developed are capable of searching for failure scenarios resulting in a violation of safety of specifications [17]–[25].

With advances in simulation-based scenario testing across domains, the key question remains — are failure scenarios detected in simulation meaningful representations of real-world behavior? The notion of the *sim-to-real gap*, the distributional shift between *real-world* and *simulated* sensor data and physics, gives rise to the phenomenon of invalid failures in simulation that lack real-world analogs [26] [27]. In order to employ simulation testing in a useful manner, there is a need to be able to separate meaningful failures with potential real-world impacts from simulation- and simulator-specific invalid failures [28]. This motivates the exploration of techniques to *validate* simulation scenario outcomes against sensor data from real-world CPS operation.

Prior work has focused on *training* CPS system components, such as AV planning and perception models, on simulated data and demonstrating robust performance on real data [29], [30]. However, the body of work focusing on *testing* simulation behavior transferability (for models that have already been trained) is more limited in scope. Physical reconstruction of these scenarios is a resource-intensive process constrained by the limitations of experimental setup. Initial approaches to handle this approach either provide no guarantees on validation outputs or are limited to the use case of static scenes. By contrast, we propose a *data-driven* approach to this problem that scales to *time series* data.

In the context of the autonomous vehicles use case, understanding (1) if real-world sensor data contains a specific realistic scenario of interest (i.e. unprotected left turn) for a specific system and (2) how the system behaves within real-world instances of the scenario is critical for safe CPS deployment. The sim-to-real gap highlights the fundamental challenges of leveraging simulations to directly predict and verify real-world system behavior [26]. However, this problem of real-world validation of simulation scenarios generalizes to many other domains including aviation, robotics, and augmented reality CPS [31]–[33]. With the rise of data-driven approaches to CPS, larger amounts of sensor data are collected across these domains [34], [35]. This leads to the question — how can we create a system that supports the (1) expression and modeling of complex, interactive multi-agent *scenarios* of interest and (2) retrieval of collected real-world data *matching* those scenarios?

Several prior works attempt to answer this question through querying approaches operating over static image or video sensor data directly. For instance, video database management systems (VDBMS) support efficient querying from multiple video sensor inputs at once. However, the domain-specific languages (DSL) used for video querying tend to be too restrictive to formally define more complex scenarios of interest. Meanwhile, though multimodal vision language model (VLM) architectures have started to demonstrate basic video understanding, many models fail to support or properly understand the nuances of 3D video sensor data input [36]. Models aligned for visual question answering (VQA) tasks lack the spatial and scenario-level reasoning required to query more complex scenarios at scale [37] and cannot provide formal guarantees on outputs [38]. In addition, VLMs demonstrate risks of hallucination and inconsistent behavior, and solutions fine-tuned to specific sensor inputs fail to generalize to other domains or use cases [39].

To address the limitations of existing solutions, we propose a sim-to-real querying system that allows for the retrieval of time series data points based on an input scenario program and real-world sensor dataset. For any CPS whose environment at any instant is a *scene*, a physical configuration of multiple agents or objects, we allow users to express distributions over scenes and agent behaviors in the SCENIC probablistic programming language [40]. At a high level, the querying algorithm automatically encodes the input probabilistic program representing a scenario as a bounded model checking (BMC) [41] problem solved with the UCLID5 modeling and verification system [42]. If the formal hierarchical encoding of the scenario could generate a specific labeled data trace, the encoded probabilistic program outputs it as a match. Formal guarantees on algorithm correctness streamline the exploration of large-scale datasets and discovery of missing and present scenarios. As sensor datasets reach unprecedented sizes [35], the need for more precisely understanding dataset contents continues to grow across domains [43], [44]. Since the algorithm queries from time series labels instead of raw sensor data, it can be extended to a wide range of use cases of unique data types such as radar, LiDAR, and RGB data. We demonstrate one application of our algorithm for validating simulation failure scenarios for a variety of dynamic autonomous vehicles tasks involving different sensor types.

**Contributions**  In this thesis, we provide:

- A novel problem formulation for querying from a labeled time series dataset using the SCENIC probabilistic programming language to construct queries for dynamic, stochastic, multi-agent scenarios and validate simulated failures against real-world examples.

- A formal proof for the matching result of a time series label, formulated as a bounded model checking problem solved by a generated UCLID5 program that provides a guarantee that queried sensor data matches a provided distribution of scenarios.

- A scalable algorithm with intuitive syntax for querying to better understand the time series sensor data contents that are present and missing in large-scale labeled datasets.

- A set of experiments demonstrating the effectiveness and scalability of the time series querying algorithm using a large-scale dataset of sensor data from real-world scenarios.

One related and concurrent work to the compilation of this thesis was the co-creation of a generative AI system for creating scenario programs from natural language descriptions in SCENIC. For more information, please refer to the publication *Generating Probabilistic Scenario Programs from Natural Language* [45].

## 1.2  Related Work

In the autonomous vehicles domain, scenario-based testing is one of the most common forms of evaluation in both simulation and real-world contexts. Disengagement reports from California Department of Motor Vehicles indicate that autonomous vehicles from companies including Waymo, Cruise, Zoox, Pony.AI traveled an average distance over 10,000 miles before encountering challenging scenarios requiring human intervention [46]. This motivates the construction of targeted driving scenarios for testing instead of waiting for higher-risk and lower-frequency scenarios to arise in everyday contexts. In order to construct testing scenarios, close studies of accident reports and naturalistic driving data can assist the creation of both simpler and more complex driving scenarios. Alternative to validating simulation failures against real-world data can involve resource-intensive physical reconstruction, scenario querying from video database management systems reliant on less flexible DSLs, more inconsistent visual question answering models, or querying-based systems built on scenario modeling languages for static data retrieval. Our approach aims to address the limitations of existing approaches by presenting a scalable, multi-agent scenario-based querying approach that supports direct simulation evaluation of formally-defined scenarios and scalable real-world data querying.

## Track Testing and Physical Scenario Reconstruction

In a similar study [47] conducted by Fremont et al., autopilot failures in simulation were verified in real-world testing through *physical* reconstruction at a designated track testing facility. The study determined the frequency with which safe and unsafe simulation runs corresponded to safe and unsafe physically reconstructed track runs, for a series of safety properties and test cases synthesized with formal methods. Despite the precision and interpretability of these results, manual validation is highly intensive in terms of labor, time, and physical resources. As a result, this approach fails to scale to the levels required to verify autonomous vehicle systems deployed on-road. In addition, the limiting assumptions most physical reconstruction setups involve, such as the assumption that AV behavior did not affect the flow of surround traffic, can constrain the types of scenarios that can be fully reconstructed and evaluated within this format. Though Kim et al. [48] automate this process for static (single-frame) scenes, our algorithm aims to extend these capabilities to dynamic multi-agent scenarios for *time series* data. We propose this solution as it most closely matches the nature of large-scale simulation testing workflows and real-world sensor datasets.

Several other projects, including the PEGASUS [49] project for autonomous driving comparison against human driving capabilities, have explored similar benchmarking and evaluation tasks through physical test construction and evaluation. Alternative approaches to leveraging direct physical reconstruction of simulation tests explore the construction of test scenarios based on the analysis of crash data or naturalistic driving data. These sources of human-created driving data can serve as inputs for human or automatic generation of driving test cases grounded in real-world human driving data. However, evaluation on these datasets in simulation can fail to provide meaningful insights into the real-world failures that would be produced in the same scenarios. Gaining meaningful insights about real-world behavior and performance would require somehow validating the failures against real-world data, likely driving this approach to be combined with another related approach or direct human evaluation. Our proposed solution not only bridges the sim-to-real gap supporting testing programs in simulation and querying with these programs, but also provides visibility into the contents of the real-world dataset and any scenarios that may be overrepresented or missing.

## Video Database Management System Retrieval

One more scalable alternative to physical reconstruction involves retrieval from large-scale video analytics systems, in order to capture real-world sensor data with semantically similar to a given test scenario. For instance, the geospatial VDBMS Apperception [50] supports the efficient retrieval of geospatial video data based on a domain-specific language. The system supports object querying in a multi-sensor system and is designed for the retrieval of time series data in the form of videos. A recent extension to Apperception, Spatialyze [51] is a VDBMS that allows users to declaratively specify and efficiently retrieve videos of interest of a range of formats. Spatialyze supports more flexible video formats and enables the retrieval of a wide variety of sensor types and camera angles. This allows for the validation of an autonomous vehicle system with the use of a single VDBMS containing all sensor types. Alternative approaches to querying from multiple videos either consider all videos independently [52]–[54] or compromise on accuracy by attempting to operate over multiple video streams [55]. Scenario retrieval from a geospatial VDBMS is a more efficient and scalable process, but still can require direct operations over large amounts of stored data per scene search. In addition, the domain-specific querying language for most VDBMS architectures do not support the full extent of usable and interpretable descriptions of formal scenarios. Our proposed approach aims to let users express queries as more expressive formal scenario programs, while querying from labeled traces to improve storage requirements, overhead costs, and generalizability to other domains.

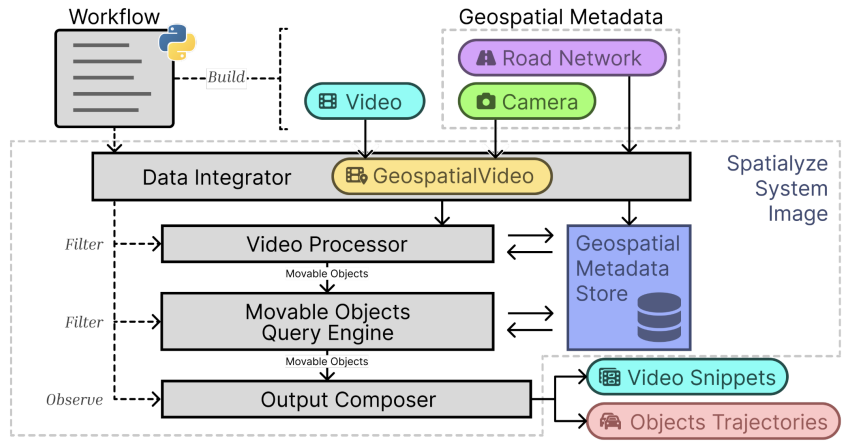

Figure 1.1: (reproduced from Kittivorawong et al. [51]) An overview of the Spatialyze VDBMS system for video data retrieval. The query is expressed in a domain specific language. The system directly operates over geospatial video data and metadata to extract video snippets and object trajectories.

## Multimodal Visual Question Answering Systems

Recent progress in space of multimodal generation and understanding has given rise to the notion of more advanced VQA [56] and VLM [57] systems. Within a VQA context, sensor data inputs can be considered and generate outputs and reasoning about the contents of different scenes in response to user input queries. For instance, the recent NuScenes-QA benchmark [58] represents the first autonomous driving VQA benchmark capable of supporting multimodal, multi-frame inputs with moving foregrounds and static backgrounds. Though VQA and VLM models demonstrate an ability to answer knowledge-based questions about videos

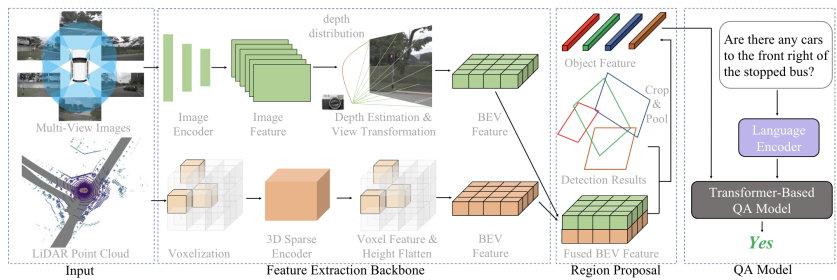

Figure 1.2: (reproduced from Qian et al. [58]) An overview of the NuScenes-QA system for visual question answering over time series autonomous driving data. User inputs take the form of natural language questions about the contents of a scene. Model responses pass through a Transformer QA model. This produces more human-interpretable, but less controllable output that can be challenging to scale to larger databases.

[59], recent works indicate these architectures demonstrate limited 3D spatial awareness and reasoning abilities [37] without specific training, fine-tuning, or dataset augmentation. In addition, many models are limited to single-frame inputs from a single consistent sensor format. The inconsistency of language model outputs, risks of hallucination, and high inference costs of attention-based language models [60] all limit the precision and scalability of VQA models for real-world scene retrieval based on specific driving scenarios. As the capabilities and consistency of VQA models continue to improve, they may eventually become more well-suited for retrieving specific scenarios from collections of real-world driving data. However, VQA models and VLMs currently demonstrate high risks of hallucination, inconsistent behavior, and spatial misunderstanding that can be detrimental for failure validation in real-world CPS contexts. By contrast, we propose an algorithm that demonstrates consistent and interpretable behavior for formally specified scenarios, which generalizes to any domain involving time series spatial sensor data without any model fine-tuning or alignment.

## Scenario Modeling Languages

Probabilistic programming languages (PPLs) provide a mechanism for guiding data generation for simulations in the general direction of scenarios of interest. In the autonomous vehicles context, the SCENIC environment modeling language [40] enables the generation and verification of targeted and meaningful driving scenarios containing autonomous vehicles in a precise, realistic, and scalable manner. Several probabilistic languages including M-SDL [61] and PROB [62] support a similar feature set more tailored towards inference tasks than scenario generation. Meanwhile, non-PPL frameworks such as Paracosm [63] model dynamic scenarios in the autonomous vehicles domain, often without the same extent of behavior models and compositional features. The SCENIC language supports concise but expressive syntax for defining spatiotemporal relationships between agents and objects [40]. As a result, it assists with the creation of more complex test scenarios in autonomous drive edge cases including occluded view situations and poor weather conditions [64].

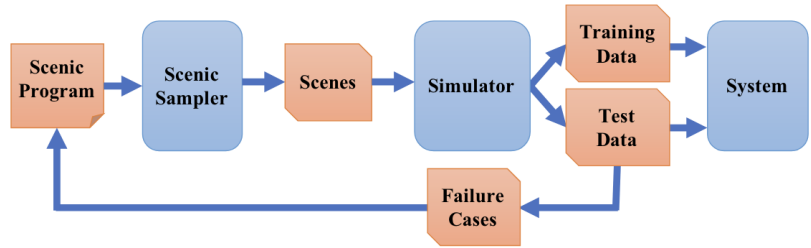

Figure 1.3: (reproduced from Fremont et al. [40]) Sample training, testing, and debugging workflow leveraging the SCENIC probabilistic language. For instance, this evaluation system can be used to detect issues with an autonomous vehicles perception or planning system.

Progress in synthetic data generation [65] (generating new data matching the real-world distribution) and domain adaptation [66] (bridging gaps between real-world and synthetic data distributions) demonstrate the opportunity to train models with synthetic data to perform tasks operating on real-world data. In the context of *training* purposes, Generative adversarial networks (GANs) [67] can transform synthetic data into more realistic data. However, our proposed solution aims to reduce the gap between simulation and real-world data in the context of *testing* autonomous vehicles and other cyber-physical systems. More specifically, we propose an algorithm to efficiently verify that pre-trained autonomous models exhibit consistent behaviors in simulation and real-world test scenario contexts. The potential use cases of the algorithm extend beyond the scope of verified autonomy to facilitate querying and understanding from existing real-world test sensor data. This creates an opportunity for users to detect missing and underrepresented scenarios in existing data.

## 1.3 Background

In order to propose a system that uses simulation programs to query real-world time series data, we first define the notion of a scenario description language, verification system, and SMT theories. Though we focus on a labeled dataset in our demonstrated use case for the autonomous driving domain, this approach can be applied to any cyber-physical system with time series data of the general format defined in the problem statement. As demonstrated in the autonomous vehicle example, any pre-trained neuro-symbolic labelling system can be used to provide labels of the form required for the querying task.

## Scenic: Scenario Description Language

SCENIC is a domain-specific language designed for specifying scenarios in simulation-based testing for cyber-physical systems [40]. It provides a high-level syntax for defining complex scenarios, including the layout of environments, positions of objects, and behaviors of agents, allowing for complex deterministic and probabilistic reconstructions of real-world scenarios. The expressive power of the SCENIC language enables users to concisely describe intricate scenarios that are crucial for testing and verifying the behavior of autonomous systems under diverse conditions. A SCENIC scenario defines a distribution over the *scenes* and the *behaviors of dynamic agents* within the scene over time. This allows for the use of more complex sampling techniques and more expressive *scenario improvisation* (random generation of concrete scenarios) to precisely match user specifications.

```
1 param weather = Uniform('sunny', 'rainy')
2 param time = Range(4,6)
3 param SAFETY_DIST = Range(3,5)
4
5 behavior FastFollowLaneBehavior():
6   try:
7     do FollowLaneBehavior()
8   interrupt when not withinDistanceToObjsInLane(self, SAFETY_DIST):
9     do AccelerateForwardBehavior()
10
11 behavior SafeFollowLaneBehavior():
12   try:
13     do FastFollowLaneBehavior() until network.intersectionAt(self)
14   interrupt when distance to pedestrian < SAFETY_DIST:
15     do BrakingBehavior()
16   do TurnRightBehavior()
17
18 ego = new Car on road, facing roadDirection
19   with behavior SafeFollowLaneBehavior()
20 pedestrian = new Pedestrian
21
22 require not (pedestrian in intersection)
```

Figure 1.4: A SCENIC program describing an ego vehicle speeding up as long as it has room ahead, which suddenly brakes for a pedestrian obstructing the road before making a right turn. The vehicle proceeds to make a right turn once it reaches the intersection.

**Scenic Program Semantics**

Fig. 1.4 demonstrates an example multi-agent SCENIC program leveraging the fragment of SCENIC supported for the querying process. To simplify the modeling process for more complex scenarios, SCENIC leverages an intuitive and interpretable syntax to define more complex relations between objects within a scene. For instance, the program in Fig. 1.4 uses the `on` specifier to uniformly generate a random point within a region. Meanwhile, the user can add a `require` statement to enforce a property across *all executions* of the scenario. The `try-interrupt` block and nested behavior definitions enable program definitions with non-deterministic outcomes depending on SCENIC random sampling and external environment variables. All executions of a SCENIC program sample a scene satisfying all user constraints defined through the entire scene, specified with `require` statements.

In this program, a vehicle is defined with behavior controlled by a nested `try-interrupt` block in SCENIC triggered by external environmental conditions for the agent. In this case, the ego vehicle attempts to follow the lane until it reaches an intersection before turning right. However, the lane following is interrupted by forward acceleration if the vehicle has enough open space. This entire set of behavior outputs is interrupted by braking if the pedestrian enters the safety distance of the vehicle at any point in time. The pedestrian is defined anywhere in the scene, while the behavior of the ego vehicle is defined according to the formal specification created by the SCENIC program. In this case, the program demonstrates several key features of the supported fragment of SCENIC supported by the querying process.

To support efficient and scalable querying, the SCENIC program encoding described above is converted into a interrupt-driven, hierarchical extended finite state machine. Please refer to Appendix A.2 for examples of this process.

**Specifying a Fragment of Scenic for Querying Task**

In order to support more complex processes involved in the querying process including hierarchical state machine code generation, program analysis, semantic transformations, and formal verification, there is a clear need to formally defined fragment of the SCENIC language.

By limiting the fragment of SCENIC to the constructs defined below, the formalization of this SCENIC language fragment maintains support for expressive conditional logic, scenario control, and reactive behaviors, supporting an extensive range of testing scenarios for autonomous systems. The exclusion of loops and iteration reflects a design choice to simplify supported scenario specifications, in light of issues such as clearly and reconciling simulation and real time. We define an Extended Backus-Naur Form (EBNF Grammar) [68] corresponding to the fragment of SCENIC for which the querying process is supported in Appendix A.1.

**Specification of Scenic Fragment**   We define the fragment of SCENIC for which our proposed algorithm is supported as follows:

1. **Operators**: All the Boolean, Orientation, Region, OrientedPoint, Temporal, and Vector operators as defined in standard SCENIC syntax.

2. **Distributions**: All distributions (Range, DiscreteRange, Normal, TruncatedNormal, Uniform, Discrete) according to their standard SCENIC distributional definitions. These distributional objects constitute one source of nondeterminism in SCENIC programs.

3. **Compound and Simple Statements**: The algorithm supports the behavior definition and try/interrupt compound statements. In terms of simple statements, global parameter definitions, terminate when/after statements, and require statements involving Boolean constraints are all supported.

4. **Objects**: The algorithm supports the definitions of all objects, in addition to the properties added by both the OrientedPoint and Object classes.

5. **Specifiers**: The algorithm supports the specifiers for any properties built into the SCENIC language.

6. **Orientation**: The algorithm supports the full extent of orientation-defining syntax built into the SCENIC language.

7. **Operators**: The algorithm supports all operators built into the SCENIC language.

8. **Dynamic Statements**: The supported dynamic statements can be found in the next section, including but not limited to sequences of do statements, nested try-interrupt blocks, and abort and terminate statements.

**Dynamic Statements Fragment Reference**  Within the classification of dynamic statements, we consider the following types of statements shown below:

(i) **Assignment Statements**: Allows variables to be assigned values, facilitating state management within scenarios.

(ii) **Try/Interrupt**: A construct for specifying behaviors that should be interrupted under certain conditions, useful for modeling reactive behaviors.

(iii) **Do and Do/Until**: Defines actions or sequences of actions that an agent should perform, with the possibility of specifying a condition under which to stop the action.

(iv) **Nested Statements**: Enables complex, nested conditional logic for statements including Try/Interrupt and If/Else statements, allowing for more detailed specifications of agent behaviors and scenario progression.

(v) **Take Actions**: Specifies immediate actions to be taken, enabling quick responses or changes in the environment or agent behavior.

(vi) **Abort**: Immediately terminates the current action or behavior, useful for modeling sudden stops or changes in behavior.

(vii) **Terminate**: Ends the scenario or a specific behavior, useful for cleanly concluding a scenario or an agent's actions.

(viii) **Require**: Specifies a dynamic hard Boolean requirement for the program as it executes.

**Excluded from Scenic Fragment**

1. **For Loops and Iteration**: Constructs for repeating actions or behaviors multiple times are not included in the currently defined fragment of the SCENIC language. This exclusion means that scenarios requiring repeated actions must find alternative means of expression, potentially limiting the ability to easily specify scenarios with repetitive or iterative behaviors.

2. **Linear Temporal Logic Require Statements**: Require statements may not be invoked with LTL expressions, and must instead be defined using the supported Boolean operators in SCENIC as defined above.

3. **Scenario Composition**: SCENIC supports advanced constructs for composition of multiple scenarios, and Do statements can be invoked to call scenarios. However, the algorithm scope will be constrained to preexisting SCENIC specifiers within the context of a single scenario program.

4. **Abort and Override Statements**: due to the temporal limitations of the UCLID5 encoding of the SCENIC program, we do not extend support to statements that may require epsilon transitions to generate an action trace for an agent at each timestep.

## UCLID5: Formal Specification Language

The sample simulation-based algorithm in Appendix A.5 captures the complexity of approaching this task without a formal specification tool. To formalize the problem, we first define an interrupt-driven, hierarchical finite state machine representation of a SCENIC program for any program contained within the fragment of SCENIC our algorithm supports. We define the *support* of a SCENIC program to consist of the set of all possible behavior outputs it can generate for a single behavior program, provided a stream of Boolean variable inputs from real-world data. A simulation-based approach to detecting if a labeled data trace is contained in the support of the SCENIC program (set of all possible traces) requires an exponentially growing [69] set of SMT constraints. This fails to properly scale to the challenge of representing nondeterministic state transitions that SCENIC supports, dramatically limiting the fragment of SCENIC that an approach of this nature would support.

As a result, we leverage the UCLID5 formal verification system [70] [71] to efficiently perform the task of bounded model checking against the formal specification of a SCENIC program. Our SCENIC to UCLID5 translator supports the programmatic modeling and construction of SCENIC IHEFSMs and trace checking against real-world labeled data traces, without constraining the supported language fragment. This includes any reset transitions, terminate conditions, and nondeterministic transitions that may appear in the IHEFSM for a specific SCENIC program. In the context of scenario querying over time series data, the translation from SCENIC to UCLID5 (using an intermediate IHEFSM representation) outlined in Appendix A.2 captures the hierarchical nature of the translation process. The counterexample-search properties of UCLID5 bounded model checking [72] allow for trace checking even in contexts with nondeterministic transitions and undefined Boolean predicates, in addition to the enforcement of specific LTL conditions including partial matches. One contribution of this work is an automated translation process from the specified fragment of SCENIC to UCLID5. Several examples of this translation process and intermediate hierarchical statecharts representations can be found in the Appendix A.6.

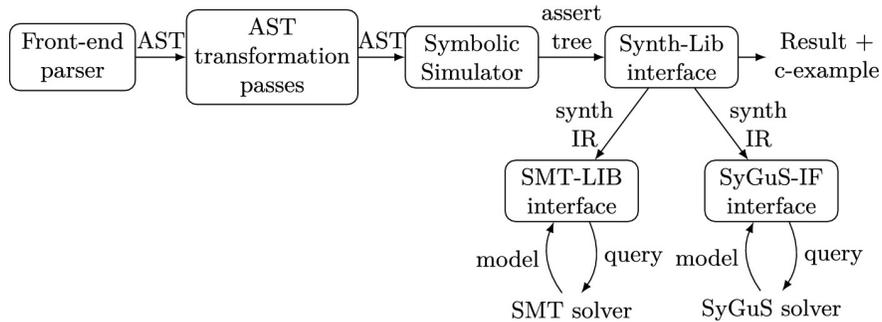

Figure 1.5: (reproduced from Polgreen et al. [70]) UCLID5 architecture diagram.

## Bounded Model Checking with Satisfiability Modulo Theories

The satisfiability problem (SAT) encompasses the notion of whether or not a satisfying assignment exists to the Boolean variables for a propositional formula. Bounded model checking (BMC) is a formal verification technique that leverages SAT solvers to check if a property holds in a finite-state model, by encoding the model and property as propositional formulas [41] [72]. Despite the theoretical hardness of checking traces against nondeterministic finite-state systems, progress in developing powerful BMC tools including CBMC [73] have enabled applications of BMC to industrial-scale problems.

In order to support problems involving arithmetic or trigonometric operations, satisfiability modulo theories (SMT) solvers extend SAT solvers by incorporating theories such as linear arithmetic. SMT solvers require a first-order logic [75] formula $\varphi$ and a background theory under which the formula is interpreted. The querying approach proposed in this thesis involves a fragment of quantifier-free nonlinear real arithmetic to support geometric and trigonometric operations, with a grammar in Appendix A.1 defining the full range of generated formulas. Although the satisfiability problem for such formulas is undecidable across real numbers, several SMT solvers (e.g., z3 with real extensions [76], dReal [77]) can return an approximate satisfying assignment or prove unsatisfiability [78].

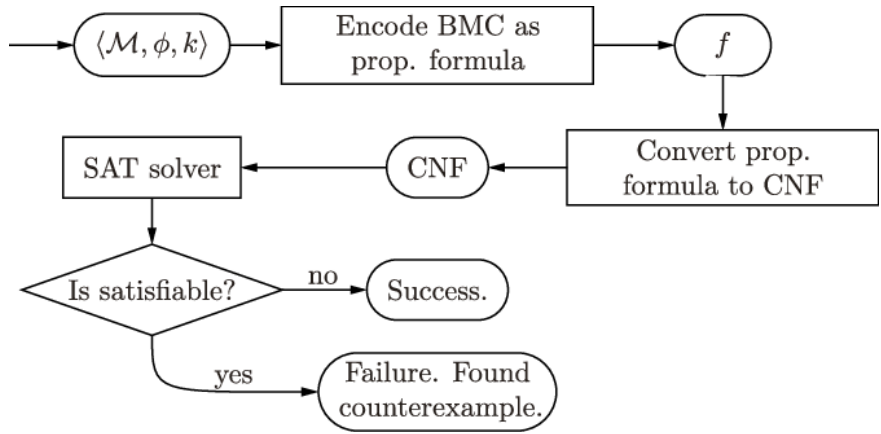

Figure 1.6: (reproduced from Andrade et al. [74]) Classical bounded model checking problem formulation. In this context, checking if SCENIC encoding produces output trace observed in NUSCENES label trace output given NUSCENES sensor data input.

In this thesis, we leverage the bounded model checking properties of UCLID5 to check if a scenario encoded as a SCENIC program matches a labeled data trace. A labeled data trace is considered a match to a SCENIC scenario if it is contained in the set of possible traces of the UCLID5 representation of the scenario. We leverage the properties of SMT solvers to handle two key tasks in considering potential matches: (1) *object correspondence* and (2) *predicate abstraction*. Object correspondence involves checking for the existence of an injective mapping from the agents defined in the SCENIC program to objects in the labeled data trace, such that the corresponding SMT formula for all mappings is satisfiable. Predicate abstraction involves replacing the predicates in the SCENIC program with Boolean or unassigned Boolean variables at each timestep, allowing for the evaluation of nondeterministic values. By encoding the SCENIC program and the data trace as an SMT formula and checking satisfiability with BMC, this ensures that user-defined variable sampling and sources of nondeterminism still result in a matching behavior trace output if one exists.

# Chapter 2

# Querying Labeled Time Series Data with Scenario Programs

## 2.1 Problem Statement

**Definition: Dynamic Multi-Agent Scenario**

The set of SCENIC fragments we allow models a scenario $P$ which represents a tuple, $(\mathcal{N}, \mathcal{T}, \{\mathcal{S}\}_{i \in \mathcal{N}}, \{\mathcal{A}\}_{i \in \mathcal{N}}, \Theta, \mathcal{I}, \mathcal{B})$. $\mathcal{N} = \{1, ..., N\}$ denotes the set of $N \geq 1$ agents defined in the program. $\mathcal{T}$ is a function which maps each agent in $\mathcal{N}$ to its agent type (e.g. vehicle, pedestrian). $\mathcal{S} := \mathcal{S}^{\mathcal{T}(1)} \times ... \times \mathcal{S}^{\mathcal{T}(N)}$ is the state space, where $\mathcal{S}^{\mathcal{T}(i)}$ denotes the state space of the agent type of agent $i \in \mathcal{N}$, which includes positions and orientations. $\mathcal{A} := \mathcal{A}^{\mathcal{T}(1)} \times ... \times \mathcal{A}^{\mathcal{T}(N)}$ is the action[1] space, where $\mathcal{A}^{\mathcal{T}(i)}$ denotes the action space of the agent type of $i \in \mathcal{N}$. The action space is finite and is defined as a set of the names of pre-defined primitive behaviors. Depending on the type of the agent, this action space can differ. If an agent type is a *vehicle*, the action space may include *'follow lane'*, *'lane change'*, etc; if *pedestrian*, then it may include *'cross road'* and *'wait'*. $\Theta$ denotes an internal state space which represents local variables instantiated and updated in the behaviors of agents as specified in $P$. $\mathcal{I}$ denotes a joint initial distribution over the state and the internal state. $\mathcal{B} : \mathcal{S} \times \Theta \rightarrow Dist(\Theta \times \mathcal{A})$ is a stochastic multi-agent policy as specified in $P$, which maps the current state $\mathcal{S}$ and internal state $\Theta$ to a joint distribution of the internal state and the action. Although the definition of a scenario resembles that of a Markov Decision Process (MDP), note that a SCENIC program $P$ does not define a state transition function and, therefore, is not an MDP.

---

[1] In the context of an action space, the notion of an action is consistent with the MDP definition of an action, rather than the SCENIC definition of an action.

## Definition: Labeled Time Series Sensor Trace

A time series labeled trace $l$, henceforth referred to as a *label trace*, denotes an observation of data over a period of time, $T$, by an *ego* agent with a number of sensors collecting data. For instance, suppose an *ego* car with RGB and LiDAR sensors drives through San Francisco to collect data for a period of time. Formally, a label trace is a finite sequence of (state, action) pairs observed by *ego*, i.e. $((s_0, a_0), ..., (s_T, a_T))$. The observed state and action, $s_t$ and $a_t$, is defined over a time varying state space, $\mathcal{S}_t := \mathcal{S}^{\mathcal{T}(1)} \times ... \times \mathcal{S}^{\mathcal{T}(N'_t)}$, and a time varying action space, $\mathcal{A}_t := \mathcal{A}^{\mathcal{T}(1)} \times ... \times \mathcal{A}^{\mathcal{T}(N'_t)}$, where the set of the number of observed agents, $\mathcal{N}'_t = \{1, ..., K_t\}$ vary over time $t$. The action space, $\mathcal{A}_t$, represents the joint actions taken by all observed agents at time step $t$. This reflects a common case in which some of the surrounding vehicles and pedestrians are observed for a duration by sensors on *ego*, but eventually are no longer observed as *ego* continues navigating through the environment. Analogs to this case naturally occur in other CPS domains such as aviation and robotics. Note that the sequence of observed states $s_0, s_1, ..., s_T$ are being updated by the ground truth dynamics of the world. For some scenario that does not define a state transition dynamics function, the sequence of states being updated by the ground truth dynamics of the world will be used as scenario inputs.

We make the following assumptions about the label trace $l$ and the scenario $P$. We assume that (1) the state space of the label trace, $\mathcal{S}'_t$ contains at least the position and the orientation of all observed agents, and (2) the state and the action spaces of the scenario and the label *for each agent type* is equivalent (e.g., all agents of type *vehicle* will share the same action space in the label and the scenario, {*follow lane, lane change, right turn, ...*}).

## Problem Formulation

Suppose we are given a time series label trace $l := ((s_0, a_0), ..., (s_{T-1}, a_{T-1}))$ of length $T$, and a SCENIC program modeling a scenario $P$ as defined above. The label trace $l$ contains a sequence of observed states and actions of $\mathcal{N}_t$ agents for $\forall t \in \{0, ..., T-1\}$, where the states are updated according to the actions and the ground truth state transition dynamics of each agent in the real world. The multi-agent policy $B$ of the scenario $P$ constrains the actions of agents according to the given states.

Let $\Pi_{P,l}$ define a set of feasible paths, $s_0 a'_1 s_1 a'_2 ... s_{T-1} \in \mathcal{S}_0 \times (\mathcal{A}_t \times \mathcal{S}_t)^*$ for $\forall t \in \{1, ..., T-1\}$, constructed using a sequence of states, $s_0, s_1, ..., s_{T-1}$, in the label $l$, where the corresponding actions $a_0, a_1, ..., a_{T-2}$ are constrained by the multi-agent behavior policy $B$ of the scenario. To formally define a path, we assume $|\mathcal{N}_t| \geq |\mathcal{N}|$ for all $T$ time steps, such that there exists a *correspondence*, $\mathcal{C} : \mathcal{N} \rightarrow \mathcal{N}_t$, i.e. a fixed injective mapping from the $\mathcal{N}$ agents in the scenario to $\mathcal{N}_t$ for all $T$ time steps. Given a sequence of states from the label, the path is feasible if there exists a sequence, $\theta_0, \theta_1, ... \theta_{T-1}$ such that $\mathcal{B}(s_t, \theta_t)(a_{t+1}) > 0$ where $\mathcal{I}_\Theta(\theta_0) > 0$ and $\mathcal{P}_\Theta(s_t, \theta_t)(\theta_{t+1}) > 0$ for all $T$ time steps. To illustrate the path generation process, given a correspondence $\mathcal{C}$, we inductively generate the path in the following way. As a base case, we compute $a_1$ using the multi-agent policy $\mathcal{B}$ in the scenario, i.e. $a_1 \in \mathcal{B}(s_0, \theta_0)$, where $\theta_0$ is a feasible internal state in the given initial internal state distribution $\mathcal{I}_\Theta$ of the scenario. Then, given $s_1$ from the label, we compute $a_1 \in \mathcal{B}(s_1, \theta_1)$ where the updated internal state $\theta_1$ is computed using the internal state transition function, $\theta_1 \in \mathcal{P}_\Theta(s_0, \theta_0)$. This way, we inductively generate the path, $\pi_{P,l}$. Let $\Pi_{P,l}$ denote the set of all paths generated by $P$ and $l$ for all possible correspondences.

**Problem Formulation P0:** Given a time series label $l$ and a SCENIC program modeling a scenario $P$, the label *matches* the scenario if (i) $\mathcal{I}_\mathcal{S}(s_0) > 0$, (ii) $|\mathcal{N}| = |\mathcal{N}_t|$, and (iii) there exists a correspondence $\mathcal{C} : \mathcal{N} \to \mathcal{N}_t$, whose mapping remains fixed for $\forall t \in \{0, ..., T-1\}$, such that the label path $l' \in \Pi_{P,l}$.

However, the above problem statement is too strict for our querying purpose, very often not matching any label traces to the scenario. This strictness is a result of the following:

(1) The label may contain more number of objects than in the scenario, i.e. $|\mathcal{N}'_t| \geq |\mathcal{N}|$ for $\forall t \in \{0, ..., T-1\}$. We may want to consider $l$ as matching $P$ even though it contains additional objects that do not have any counterpart in $P$. For instance, we may want a program a scenario for "a pedestrian crossing a road" to match whenever such a pedestrian exists, even if there is a second pedestrian in the label.

(2) Requiring the full duration, $t \in \{0, ..., T\}$, of the label trace to match the scenario may be too strict, especially as the duration increases. Thus, we introduce a new parameter $m$ to define the minimum consecutive time duration for the label trace to match the scenario.

**Problem Formulation P1:** Suppose a time series label $l$ of length $T$, a scenario $P$, and a minimum time duration $m$ are provided. Then, the label *matches* the scenario if there exists $j, k \in \{0, ..., T-1\}$, where $k - j > m$, such that $\mathcal{I}_\mathcal{S}(s_j) > 0$, $|\mathcal{N}_t| \geq |\mathcal{N}|$, and there exists a fixed injective mapping $\mathcal{C} : \mathcal{N} \to \mathcal{N}_t$ for $\forall t \in \{j, ..., k\}$, such that there exists a path $\pi_{P,l} \in \Pi_{P,l}$ which is equivalent to the label path, $l'$, for the consecutive time steps, $j$ to $k$.

## 2.2 Methodology

Given a time series label and a SCENIC program, our approach translates the program to an SMT formula that is satisfied if and only if the time series label constitutes a match to the program as previously defined for a window of $m$ timesteps. Fig. 2.1 depicts the architecture for the process of proposing a candidate object correspondence described in Algorithm 1.

### Checking All Object Correspondences

In order to determine if each agent defined in a scenario $P$ is represented in a labeled data point $l$, we define an integer SMT formula encoding the object correspondence problem. The SMT formula searches for an injective function mapping from the set of SCENIC agents $\mathcal{N}$ to the set of labeled data objects $\mathcal{N}'$, such that each agent $i$ has a corresponding object $\mathcal{C}(i)$ in the labeled data. A key assumption of this process is that $|\mathcal{N}| \leq |\mathcal{N}_t|$, guaranteeing an injective mapping from the scenario program to the labeled data. This assumption is presented since many objects may enter and exit the field of view in even a small timeframe

Figure 2.1: Algorithm 1 overview from inputs $P$ and $l$ to outputs (match/no match). At each iteration, the SMT solver proposes a new candidate object correspondence. The UCLID5 IHEFSM evaluates the object correspondence for the label trace available and finds satisfying values if any exist to make $l \in L$ for the specified threshold $m$. If none exists, the correspondence is marked as a failure and the process repeats.
(1) SCENIC converted into intermediate IHEFSM representation. (2) Predicate abstraction on program ready to accept inputs from labeled data. (3) UCLID5 executable IHEFSM is generated based on extraced conditions and intermediate representation. (4) Object correspondence is initiatlized. (5) Data is augmented with labels from computer vision labeling system. (6) Labeled behavior traces generated and indexed by object class for object correspondence. (7) UCLID5 program receives next object correspondence proposal and corresponding Boolean streams and labeled traces as inputs. (8) BMC on UCLID5 program checks if trace could be generated by IHEFSM representation of SCENIC scenario.

**Algorithm 1** Determining if SCENIC program $P$ matches a time series label $l$
---
**Input**: SCENIC program $P$, a time series label $l$, and a library of behaviors $B$, and a minimum time duration $m$
**Output**: Does $l$ match $P$? (True / False)
1: $AST \leftarrow Compile(P)$ // get abstract syntax tree (AST)
2: $\phi \leftarrow InitializeCorrespondenceSMTConstraints(l, AST, m)$ // Algorithm 2
3: **while** $SMTSolver(\phi)$ has a solution **do**
4:    $corr \leftarrow SMTSolver(\phi)$ // a feasible object correspondence $corr$
5:    $isMatch, \phi' \leftarrow MembershipQuery(l, AST, corr, m, B)$ // Algorithm 3
6:    **if** $isMatch$ is $True$ **then**
7:      **return** $True$
8:    **else**
9:      $\phi \leftarrow \phi \wedge \phi'$
10: **return** $False$

---

of *ego* sensor data collection. If the condition asserting an existence of a match $\mathcal{C}(i)$ is violated for *any* of the agents $i$ defined in the multi-agent scenario under a specific candidate correspondence, the candidate correspondence fails and a new correspondence is tested. This process repeats and the SMT solver iteratively proposes new candidate matches until a candidate either matches (outputs match) or makes the system of possible assignments from $l$ to $L$ unsatisfiable (outputs no match). Algorithm 2 formally defines this process for a specific scenario $P$ and labeled data element $l$.

## Modular Evaluation of Candidate Correspondence

In order to evaluate a specific correspondence, the following algorithm converts from the SCENIC program $P$ and labeled data $l$ to check if $l \in L$ for program $P$.

**Predicate Abstraction** The predicate abstraction process outlined in Algorithm 3 converts all atomic Boolean variables responsible for the control of execution of the SCENIC program into an abstracted representation. This allows the values in $l$ that may affect the execution of the conditions and corresponding SCENIC behavior definitions to be computed as a preprocessing step before generating UCLID5 code. This dramatically reduces the complexity, while allowing more complex SMT solvers with support for reals including z3 to be used for the task of predicate abstraction. Fig. 2.2 depicts the postprocessed SCENIC program after the predicate abstraction process. This postprocessed program is converted into an interrupt-driven, hierarchical extended finite state machine that takes in streams of abstracted predicates as inputs. The predicate abstraction process of pre-computing Boolean conditional variables reduces overhead for the model checker and simplifies the introduction of nondeterminism into the supported SCENIC fragment.

**Algorithm 2** Identify SMT Correspondence

**Input**: a time series label $l$, an abstract syntax tree (AST), and minimum time duration $m$
**Output**: (1) [True/False] is there a feasible mapping between agents in scenario and label? (2) SMT formula encoding the constraints on the mapping between the agents specified in the AST and the agents in the label $l$

1: $labelAgents \leftarrow$ dictionary // {key: agent class, value: an empty list}
2: $scenarioAgents \leftarrow$ dictionary // {key: agent, value: an empty list}
3: $\phi \leftarrow$ True // initialize a SMT formula
4: **for each** agent $i$ in the label $l$ **do**
5:   **if** (length of agent $i$'s trajectory) $\geq m$ **then**
6:     Add agent $i$ to $labelAgents[i$'s class type$]$
7: **for each** agent $j$ in $AST$ **do**
8:   behaviorSet $\leftarrow$ find a set of j's feasible behaviors
9:   **for each** agent $k$ in $labelAgents[$agent j's class type$]$ **do**
10:     behaviorSet' $\leftarrow$ find a set of k's feasible behaviors
11:     **if** behaviorSet' $\subseteq$ behaviorSet **then**
12:       Add agent $k$ to the list, scenicAgents[j]
13:   **if** scenicAgents[j] is empty list **then**
14:     **return** False, None // No feasible counterpart agent in the label
15: **for each** agent $i$ in $scenarioAgents$'s keys **do**
16:   $\phi \leftarrow \phi \land$ (SMT formula encoding agent $i$ can be mapped to any agents in scenarioAgents[i])
17: $\phi \leftarrow \phi \land$ (SMT formula encoding each scenic agent must be mapped to a unique agent in the label)
18: **return** True, $\phi$

**Algorithm 3** Pre-Compute Transition Conditions

**Input**: object correspondence ($corr$), a time series label ($l$), abstract syntax tree of a Scenic program ($AST$), and scenic agent ($i$)
**Output**: A dictionary of each transition condition in agent $i$'s behavior to a sequence of Boolean evaluations of the condition using $l$ for all time steps in $l$

1: conditions $\leftarrow$ dictionary // {key: condition, value: an empty list}
2: $AST_i \leftarrow ParseBehavior(AST, i)$ // scenic agent i's behavior AST
3: **for each** condition $c$ in $AST_i$ **do**
4:   $\psi \leftarrow$ SMT translation of the condition
5:   **for each** time $t$ in $l$ **do**
6:     $\psi \leftarrow \psi \land$ (SMT encodings of referred state values of agents in $l$ according to $corr$ at time $t$)
7:     $boolean_c^t \leftarrow SMTSolver(\psi)$ // evaluate the condition $c$ at time $t$
8:     Add $boolean_c^t$ to the list conditions[c]
9: **return** conditions

```
11 behavior SafeFollowLaneBehavior():
12   try:
13     do FastFollowLaneBehavior() until cond_until_2_0
14   interrupt when cond_interrupt_1_1:
15     do BrakingBehavior()
16   do TurnRightBehavior()
17
18 ego = new Car on road, facing roadDirection
19   with behavior SafeFollowLaneBehavior()
20 pedestrian = new Pedestrian
21
22 require not (pedestrian in intersection)
```

Figure 2.2: A SCENIC program after the predicate abstraction step, replacing more complex `do until` and `interrupt` conditions with standard Booleans that can be filled in by the predicate abstraction module.

**Conversion to UCLID5 IHEFSM** For the specified fragment of SCENIC, each class of statement is automatically parsed into an interrupt-driven, hierarchical extended state machine. The IHEFSM outputs a set of atomic behaviors at each statement when provided with abstracted predicate inputs. This intermediate IHEFSM representation is automatically translated into the Statecharts language from UCLID5, with supported renderings automatically generated from PlantUML included in Fig. A.1. From a hierarchical representation of the intermediate statecharts, each UCLID5 module is defined in a hierarchical manner, defining instances of child modules as required until all base cases are reached in the SCENIC to UCLID5 translation process. The final state representation of the SCENIC program in UCLID5 generates a behavior trace at each timestep, accepting condition inputs

---

**Algorithm 4** SCENIC to UCLID5 IHEFSM Conversion
---
**Input**: full SCENIC program ($P$)
**Output**: symbolic IHEFSM representation of the SCENIC program and UCLID5 hierarchical program encoding for each defined behavior
1: $AST \leftarrow Compile(P)$ // generate abstract syntax tree ($AST$) for scenic program $P$ or output SCENIC syntax error if $P$ is malformed
2: $behavior\_name\_to\_uclid \leftarrow$ dictionary // {key: behavior name, value: behavior IHEFSM}
3: **for** $behavior\_name, behavior\_definition$ in $AST$ **do**
4:     $behavior\_name\_to\_uclid[behavior\_name] \leftarrow ParseName(behavior\_definition)$
5: **return** $behavior\_name\_to\_uclid$ // each behavior name key maps to a value containing the corresponding UCLID5 IHEFSM code from abstracted predicates. Please refer to Algorithm 6 (*ParseName*), Algorithm 7 (*ParseSequence*), and Algorithm 8 (*ParseStatement*) in Appendix A.2 for a detailed description of the parsing of a specific SCENIC behavior into its corresponding UCLID5 IHEFSM.

**Bounded Model Checking** For the translated UCLID5 IHEFSM and generated streams of predicates based on SMT solver outputs, the UCLID5 module input generator is created based on labels for $l$ and condition values from the predicate abstraction preprocessing step. Algorithm 4 highlights the process of generating the UCLID5 program from the SCENIC abstract syntax tree, outlined in greater detail in Appendix A.2. The condition values serve as an input to the UCLID5 behavior module, representing the behavior definition for an agent as an interrupt-driven, hierarchical, extended finie state machine that is automatically generated from SCENIC code for the defined fragment. Simulating the state machine encoding of the SCENIC program would result in an exponential increase in overhead, failing to efficiently scale for more complex queries. However, the bounded model checking functionality UCLID5 supports as a verification tool allows for the efficient verification of matching and failing traces $l$ against a UCLID5 encoding of $P$. This scalability even holds for more complex assertions, such as the notion of partial matches of length $m$ as defined previously.

---

**Algorithm 5** Membership Query

---

**Input**: a time series label ($l$), abstract syntax tree ($AST$), an object correspondence ($corr$), minimum time duration ($m$), and a library of primitive behaviors ($B$)
**Output**:
(1) [True/False] is $l_{out}$ an element of the set of traces generated by *IHEFSM* with $l_{in}$?
(2) if False, output a SMT formula encoding the object correspondence which result in a query mismatch (otherwise, output None)

1: $dependency \leftarrow AgentDependency(AST)$ // a list of tuples of dependent agents
2: $\phi' \leftarrow True$
3: **for each** *tuple* in *dependency* **do**
4:    **for each** SCENIC agent $i$ in *tuple* **do**
5:      $l_{in}^i \leftarrow PreComputeTransitionConditions(corr, l, i, AST)$//Alg. 4
6:      $l_{out}^i \leftarrow GenerateBehaviorTrace(l, B, i)$
7:      $\psi \leftarrow TranslateToSMT(i, AST, l_{in}^i, l_{out}^i)$ // Encode AST as Interrupt-driven Hierarchical Extended Finite State Machine (IHEFSM) in SMT via UCLID
8:      **if** $SMTSolver(\psi)$ is True **then**
9:         **return** (True, None)
10:     **else**
11:         $\phi' \leftarrow \phi' \wedge$ (SMT encoding the negation of the object correspondence of all agents in *tuple* according to *corr*)
12:         **break** // out of the inner for-loop
13: **return** (*False*, $\phi'$)

**SMT Formula Updating** If the candidate correspondence fails to output a match, the SMT formula referenced in Algorithm 2 is updated to include a constraint $(i \neq \mathcal{C}(i))$ on the object $i$ that resulted in the object correspondence algorithm failure. Once the SMT formula becomes unsatisfiable, no possible object correspondences exist for $L$ that contain the trace $l$. In this case, the algorithm will terminate. Until then, the algorithm will continue to output new possible object correspondences until the process terminates with either a match or failure.

## 2.3 Experiments

As we propose a solution to the time series data querying problem for the use case of validating failures in simulation scenarios against real-world data, we observe that the query itself is the most time and resource-intensive part of the validation process. However, unlike the case of static sensor data retrieval, the time series case requires more advanced analysis of ground truth captions or sensor data to assess the performance of the querying approach. To evaluate the relevance of algorithm outputs, we consider two key questions:

(1) Given a real labeled dataset of time series data and an input SCENIC program, does the algorithm effectively retrieve relevant data matching the input program?

(2) What scaling properties does the algorithm demonstrate, in terms of scenario complexity, agents present, and time series data length?

The first question highlights the *performance* of the queried solutions by SCENIC experts in comparison to human-selected ground truths, evaluated in the *Efficacy Experiment*. The experiment construction aims to demonstrate that the notion of formal time series data querying we define and the definition of a match we output is relevant and efficient in comparison to human querying. The second experiment explores the feasibility of this approach for scenarios with more complex definitions, more agents present, and longer time series data point lengths. This experimental methodology draws from the experimental setup of Kim et al. [48] in the formal retrieval of static sensor data.

Both experiment designs use the NUSCENES autonomous driving dataset augmented with behavior labels from a vision-based behavior classification system for time series driving sensor data. In addition, the video querying system outputs generated for all experiments involved the previously defined procedure of automatically translating the SCENIC program for querying into a hierarchical, interrupt-driving finite state machine format as defined in the UCLID5 example program contained in Appendix A.3.

### Experiment I: Performance Experiment

**Setup** No algorithmic baseline exists for the task of querying time series data from a formal scenario description, and no autonomous driving datasets provided comprehensive formal scenario descriptions that could be used for algorithm evaluation. As a result, we define five scenarios of interest, ensuring a concrete example of each one appears at least once in the nuScenes dataset. We ask 3 Scenic experts to query matching time series data points for each of the 5 scenarios using the software. In order to acquire the most accurate subsets, we define the test set as the intersection of each expert's system-queried test set. We manually evaluate the correctness of each queried result to deliver accuracy metrics for each of the retrieved scenes and expert users.

**Scenarios** We define five scenarios in a range of realistic traffic situations at differing risk levels and frequencies within the dataset. We ensure that each of the five scenarios defined exists in the labeled dataset nuScenes in some capacity, and we provide natural language descriptions and corresponding Scenic encodings for three examples in the section below.

(1) Jaywalking pedestrian triggered sudden braking

(2) Yielded to another vehicle while making right turn

(3) Activated braking in response to braking leading vehicle

**Data** In order to understand the current results of their query, we provide an indication of number of results found as the user queries. To allow expert Scenic users to iterate on their query, we provide RGB video thumbnails captured from the driver's view (front camera) of the nuScenes data collection vehicle for each retrieved data element. The experts may make any modifications to their queries until the time allotment for querying this scene runs out. At this point, the Scenic scenario they have used to query a scene by natural language description constitutes their final query for the scene. The effectiveness of their query is evaluated based on the precision and recall metrics of their final retrieved labeled traces compared against human-annotated ground truths. We provide a fully worked example of a simpler form of this evaluation in the results section that follows. For this experiment, a Scenic expert is shown the driver view of a specific labeled trace $l$ from the nuScenes dataset, which aligns with one of the five evaluation descriptions. The Scenic expert is instructed to construct a scenario query to write a query containing this scenario and similar ones. All labeled traces contain sensor data collected at a rate of 2 Hz over 20 seconds, the default specifications of the nuScenes dataset. Experts query over labeled traces for match length $m=10$.

**Results** The results of a preliminary experiment indicate Scenic experts can effectively leverage the algorithm to query for a scene set containing the desired labeled trace $l$ with scenario program $P$. This serves as an effective demonstration of the precision and flexibility of the algorithm to capture the underlying distribution of a user-specified scenario. We include one full example expert query of a desired trace $l$ with $P$ for *Scenario 1*.

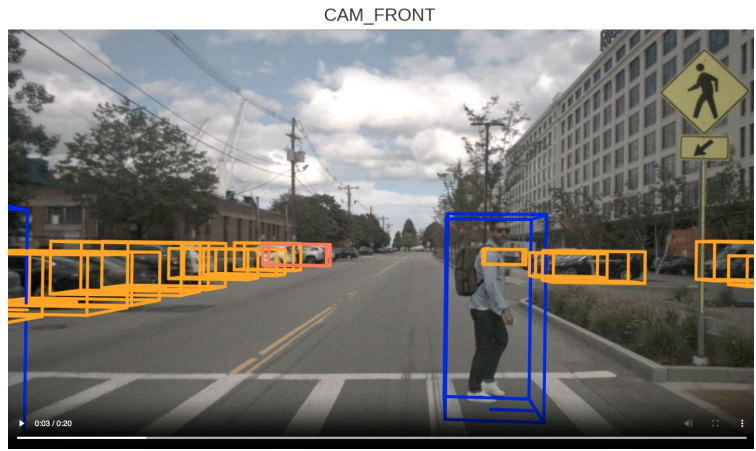

Figure 2.3: One frame of the front camera time series data from retrieved labeled trace $l$ matching expert query program $P$. Blue bounding boxes denote pedestrians and orange bounding boxes denote vehicles. This frame captures the vehicle as it brakes in response to the pedestrian. At this point, it continues to conduct `BrakingBehavior` until the pedestrian leaves the safety distance, allowing `FollowLaneBehavior` to resume.

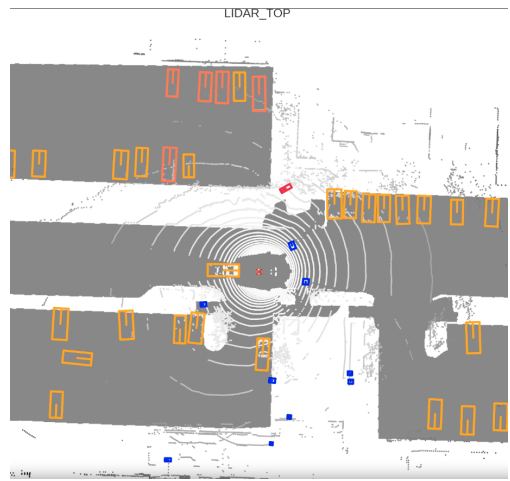

Figure 2.4: One frame of the LiDAR time series data retrieved from $l$ matching scenario query program $P$. Blue bounding boxes denote pedestrians, orange bounding boxes denote vehicles, and the red bounding box denotes a cyclist. Each line displays the heading of the corresponding bounding box. This frame captures the vehicle as it brakes in response to the pedestrian to the front and right of the vehicle (from a driver perspective). It conducts `BrakingBehavior` until the pedestrian leaves the safety distance.

**Retrieved Sensor Data**   The end-to-end querying process retrieves a successful match for the expert constructed query program $P$, as indicated by the retrieved LiDAR and RGB sensor output excerpted from the desired target time series labeled trace $l$ in Figs. 2.3, 2.4.

**Natural Language Description**   *Scenario 1* as defined above is a multi-agent stochastic scenario involving a pedestrian and a vehicle. If the pedestrian jaywalks in a manner that blocks the `ego` vehicle, the vehicle should brake in response to this until it is safe to resume the original path following the lane.

**Expert-Specified Scenic Program**   The expert-written SCENIC program provides a description of a vehicle suddenly braking for a jaywalking pedestrian. The pedestrian must violate the safety distance of the vehicle, a random parameter uniformly sampled from $[1, 10]$. This interrupts the `FollowLaneBehavior` and triggers the `BrakingBehavior` until the interrupt handler completes.

```
behavior EgoBehavior():
    try:
        do FollowLaneBehavior()
    interrupt when (distance from self to ped) < Range(1,10):
        do BrakingBehavior()

ego = new Car with behavior EgoBehavior()
ped = new Pedestrian
```

**Set of Possible Correspondences**   The SMT solver considers the set of all possible correspondences for the scene description. In this case, the role of the `ego` vehicle is fixed.

```
possible_correspondence: {'ego':['ego_0'],'ped': ['human.pedestrian.adult_0',
'human.pedestrian.adult_1', 'human.pedestrian.adult_2', 'human.pedestrian.adult_3',
'human.pedestrian.adult_4', 'human.pedestrian.adult_5', 'human.pedestrian.adult_6',
'human.pedestrian.adult_7', 'human.pedestrian.adult_8', 'human.pedestrian.adult_9',
'human.pedestrian.adult_11', 'human.pedestrian.adult_12',
'human.pedestrian.adult_15', 'human.pedestrian.adult_16',
'human.pedestrian.adult_17']}
```

**Labeled Trace $l$ for One Object**   The labeled trace below includes the input time series data processed by the predicate abstraction and precomputation process. The *xs*, *ys*, *poses*, and *angles* correspond to the x-position $X$, y-position $Y$, 3D quaternion $\mathcal{H}$, and 2D yaw angle $\Theta$. The *type* and *desc* classes contain information about the object type and description. The timesteps *ts* of the labeled trace represent up to 20 seconds of data collected at 2 Hz. Each object in a labeled trace $l$ has a similar time series set of data points to the data below. Observe that the *pedestrian* is only present for a subset of the full labeled sensor trace $l$.

```
{'xs': [2274.997, 2275.048, 2275.062, 2275.076, 2275.089, 2275.117, 2275.131,
2275.189, 2275.18, 2275.09, 2275.001, 2274.911, 2274.822, None, None, None,
None, None, None, None, None, None, None, None, None, None, None, None, None,
None, None, None, None, None, None, None, None, None, None, None], 'ys':
[849.647, 849.646, 849.648, 849.65, 849.652, 849.655, 849.657, 849.656, 849.657,
849.658, 849.659, 849.66, 849.661, None, None, None, None, None, None, None,
None, None, None, None, None, None, None, None, None, None, None, None, None,
None, None, None, None, None, None, None], 'ts': [0.0, 0.5, 1.0, 1.5, 2.0, 2.5,
3.0, 3.5, 4.0, 4.5, 5.0, 5.5, 6.0, 6.5, 7.0, 7.5, 8.0, 8.5, 9.0, 9.5, 10.0, 10.5,
11.0, 11.5, 12.0, 12.5, 13.0, 13.5, 14.0, 14.5, 15.0, 15.5, 16.0, 16.5, 17.0,
17.5, 18.0, 18.5, 19.0, 19.5], 'poses': [[0.999981235472795, 0.0, 0.0,
-0.006126067441877257], [0.999981235472795, 0.0, 0.0, -0.006126067441877257],
[0.999981235472795, 0.0, 0.0, -0.006126067441877257], [0.999981235472795, 0.0,
0.0, -0.006126067441877257], [0.999981235472795, 0.0, 0.0, -0.006126067441877257],
[0.999981235472795, 0.0, 0.0, -0.006126067441877257], [0.999981235472795, 0.0,
0.0, -0.006126067441877257], [0.999981235472795, 0.0, 0.0, -0.006126067441877257],
[0.999981235472795, 0.0, 0.0, -0.006126067441877257], [0.999981235472795, 0.0,
0.0, -0.006126067441877257], [0.999981235472795, 0.0, 0.0, -0.006126067441877257],
[0.999981235472795, 0.0, 0.0, -0.006126067441877257], [0.999981235472795, 0.0,
0.0, -0.006126067441877257], None, None, None, None, None, None, None, None, None,
None, None, None, None, None, None, None, None, None, None, None, None, None,
None, None, None, None, None], 'angles': [-0.7020000097690631,
-0.7020000097690631, -0.7020000097690631, -0.7020000097690631,
-0.7020000097690631, -0.7020000097690631, -0.7020000097690631,
-0.7020000097690631, -0.7020000097690631, -0.7020000097690631,
-0.7020000097690631, -0.7020000097690631, -0.7020000097690631, None, None, None,
None, None, None, None, None, None, None, None, None, None, None, None, None,
None, None, None, None, None, None, None, None, None, None, None], 'type':
'human.pedestrian.adult', 'desc': 'Adult subcategory.'}
```

**Boolean Condition Formula** The predicate abstraction process operates on the above object time series data for each object present, in order to generate valid predicate outputs for all Boolean variables controlling the flow of execution for the SCENIC program above. In this case, the program is governed by a single condition labeled `nusc_cond_interrupt_1_1` corresponding to the expert-written input `(distance from self to ped) < Range(1,10)`.

```
(set-logic QF_NRAT)
(declare-fun self_position_x () Real)
(declare-fun self_position_y () Real)
(declare-fun self_angle () Real)
(declare-fun ped_position_x () Real)
(declare-fun ped_position_y () Real)
(declare-fun ped_angle () Real)
(declare-fun range0 () Real)
```

```
(assert (and (<= 1 range0) (<= range0 10)))
(assert (< (+ (* (- self_position_x ped_position_x) (- self_position_x ped_position_x)) (* (-
(assert (= self_position_x 2271.70))
(assert (= self_position_y 877.18))
(assert (= self_angle 2.95))
(assert (= ped_position_x 2274.82))
(assert (= ped_position_y 849.66))
(assert (= ped_angle (- 0.70)))
(check-sat)
(get-model)
(exit)
```

**Bounded Model Checking**   The output below constitutes one example sequence of predicates input to the UCLID5 IHEFSM program included in Appendix A.3, for which the output matches the desired *nusc_trace* for at minimum $m=10$ timesteps. Based on this output, the *ego* and *pedestrian* create a matching correspondence to the program provided. As a result, $l$ is considered a match to $P$ for the satisfying assignment of objects returned by the SMT solver.

```
PREDICATE ABSTRACTION: {'nusc_cond_interrupt_1_1': [True, True, True,
True, True, True, True, True, True, True, True, False, False,
None, None, None, None, None, None, None, None, None, None, None, None,
None, None, None, None, None, None, None, None, None, None, None,
None, None, None], 'nusc_trace': ['(init)', '(init)', 'BRAKE', 'BRAKE',
'BRAKE', 'BRAKE', 'BRAKE', 'BRAKE', 'BRAKE', 'BRAKE', 'FOLLOW_LANE',
'FOLLOW_LANE', 'FOLLOW_LANE', 'ACCELERATE', 'ACCELERATE', 'ACCELERATE',
'ACCELERATE', 'ACCELERATE', 'FOLLOW_LANE', 'FOLLOW_LANE', 'FOLLOW_LANE',
'FOLLOW_LANE', 'FOLLOW_LANE', 'FOLLOW_LANE', 'FOLLOW_LANE', 'FOLLOW_LANE',
'FOLLOW_LANE', 'FOLLOW_LANE', 'FOLLOW_LANE', 'FOLLOW_LANE', 'FOLLOW_LANE',
'FOLLOW_LANE', 'FOLLOW_LANE', 'FOLLOW_LANE', 'FOLLOW_LANE', 'FOLLOW_LANE',
'FOLLOW_LANE', 'FOLLOW_LANE', '(end)', '(end)']}
```

## Experiment II: Efficiency Experiment

**Setup**   We define three scalability measures for evaluation in the efficiency experiment: scenario complexity, agent quantity, and time series data length. For each of the evaluation measures, we define a set of query programs to test the scalability of the algorithm. In terms of scenario complexity, we define four behavior definitions do until (N), do (N), try-(N)terrupt, and (N)ested-try-interrupt of the following forms. The full programs for each set of statement and time series data scalability evaluations can be found in Appendix A.4.

For each of the behavior definitions provided below, we define two independent variables consistent across the four evaluated benchmark scenarios. The *timestep* parameter denotes the number of timesteps in each labeled trace. The parameter $N$ corresponds to a notion of complexity defined for each scenario format presented below. In order to provide consistent and reproducible timing outputs, we generate $k=10$ randomized Boolean input traces and feasible randomized output traces and assert a minimum match of length $m$ for each query. This prevents queries from terminating in less than worst-case runtime and allows for the construction of runtime distributions for each statement type, complexity parameter $N$, and number of timesteps $T$. We also set a hard maximum evaluation time $e_{max}$=60 seconds for any evaluation component to improve evaluation efficiency on single-chip systems.

**Scenario Format: do until (N)** The `do until (N)` behavior definition tests the scalability of the simple do-until statement representation in UCLID5 and its ability to handle large sequences of statements driven by labeled data and user inputs.

```
behavior do_Ntil():
  do until ..
  .. repeat N times ..
```

**Scenario Format: do (N)** The `do (N)` behavior definition tests the scalability of the simple do block statement in UCLID5 and its ability to handle large sequences of statements driven by nondeterminism. To improve querying system usability, the `do` statement allows for any satisfying (uninterpreted) Boolean to fill the place of the `until` block in a traditional `do until` statement.

```
behavior do_N():
  do ..
  .. repeat N times ..
```

**Scenario Format: try-(N)terrupt** The `try-(N)terrupt` behavior definition tests the scalability of the try-interrupt block representation in UCLID5 to handling a large amount of nondeterminstic transitions or transitions driven by labeled data and user inputs.

```
behavior try_Nterrupt():
  try:
    ..
  interrupt when ..
    ..
  .. repeat N times ..
```

**Scenario Format: (N)ested-try-interrupt** The `(N)ested-try-interrupt` behavior definition tests the scalability of the nesting of multiple try-interrupt block representations in UCLID5 to support both nondeterminstic transitions and transitions driven by labeled data and user inputs at scale.

```
behavior Nested_try_interrupt():
  try:
    try:
      .. repeat N times ..
  interrupt:
    ..
```

For both behavior definition formats, we plot the average runtimes for each of these as a function of N, with a maximum timeout. In order to ensure queries do not terminate early, affecting the runtime of the scaling tests we construct randomized Boolean input traces and output traces sampled from the set of all possible traces of the `do until (N)`, `do (N)`, `try-(N)terrupt`, and `(N)ested-try-interrupt` scenario programs. In this manner, we attempt to approximate the worst-case runtime for the querying system in each of these scaling tests for program complexity. We extend this approach to testing number of agents and number of timesteps, independently plotting the runtime as a function of number of agents involved in the querying process, and as a function of the temporal length of labeled data points. In order to evaluate the scaling properties of the querying process with respect to the number of agents, we capture the average number of scenario and real-world agents, before combining data from multiple independent data points to generate larger scenes. We scale the number of agents defined in our test scenario according to the same proportionality, plotting the results against the number of objects contained within a labeled data point and number of agents defined in a scenario.

We observe that the evaluation framework presented benefits from the usage of a worst-case implementation of the UCLID5 matching process. This provides more reasonable and predictable comparisons across values of $N$ and $T$ for different statements and reduces the risks of early stopping due to unreasonably generated random Boolean traces. In order to guarantee the entire trace is processed, we propose the notion of a maximum contiguous match threshold. Within this definition of a match, for some user-specified threshold $m$, the length of a match must be larger than $m$. We leverage this definition of a match in all evaluations below.

**Results Overview**  We observe the following general trends when comparing the different SCENIC statements within our supported fragment. Based on our scalability evaluations, sequences of `do until` and sequences of `do` statements appear to run in approximately constant time regardless of sequence length. Though this is expected for `do until` statements, the ability of UCLID5 to match this performance for `do` statements is quite impressive. This is because UCLID5 must consider the additional challenges of exploring all possible `until` condition Boolean assignments across all timesteps. These results demonstrate the scalability of UCLID5-based match evaluation to handling basic user-specified nondeterminism.

**Results per Statement**  We observe the following more specific trends when comparing the different SCENIC statements within our supported fragment. We mark any evaluations that exceed the maximum evaluation runtime $e_{max}$=60 seconds with – for readability.

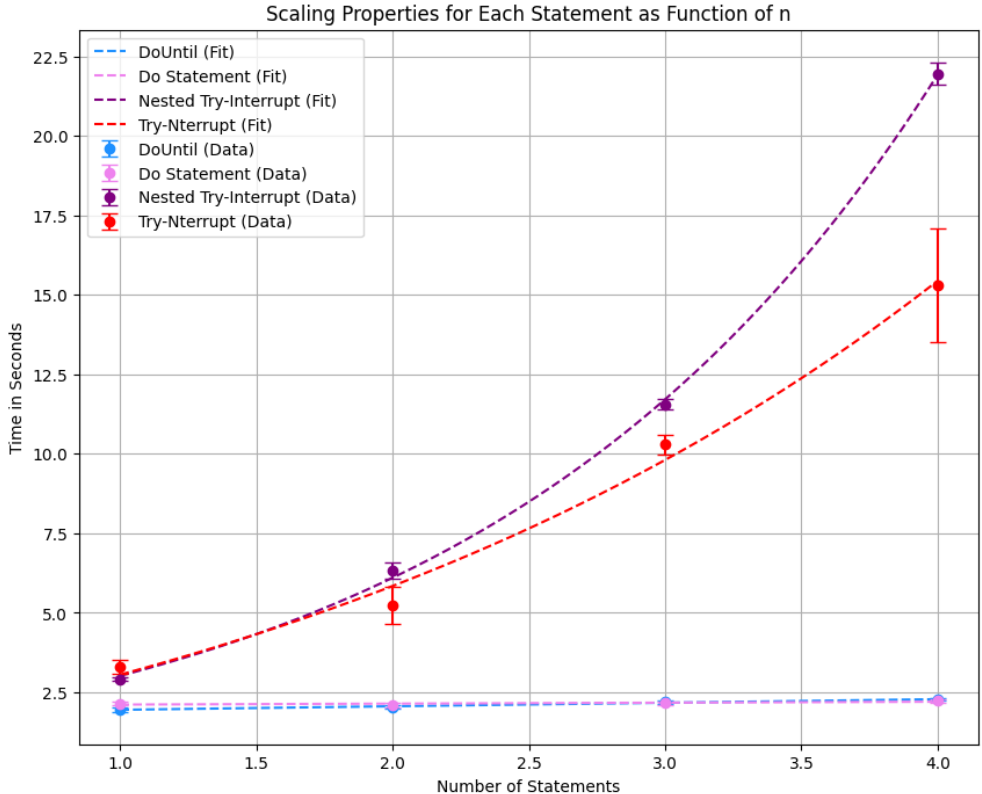

Figure 2.5: Scaling properties for each SCENIC statement as function of $N$. We observe that the *do until (N)* and *do (N)* demonstrate constant runtimes regardless of sequence length. Meanwhile, the *try-(N)terrupt* and *(N)ested-try-interrupt* statements demonstrate gently exponential scaling properties. The non-zero runtime for $N=0$ complexity captures the fixed overhead costs associated with creating input and termination handling modules in even the simplest scenario case. For this experiment, we set $k=10$, $T=10$, and $e_{max}=60$.

Based on the evaluations, scaling to much longer labeled trace lengths appears to impact overhead the most. However, we observe that many commercial AV motion forecasting datasets tend to leverage shorter scenes of 10-20 seconds, e.g. Waymax (9 seconds) [14], Argoverse 2 (11 seconds) [79], [80], NUSCENES (20 seconds) [81]. In cases with significantly longer traces, we propose several properties of the algorithm that allow for further scalability improvements.

## 2.4 Discussion

The analysis of the Experiment II results substantiate our claims about the scalability of the system to user-specified nondeterminism, the full defined SCENIC fragment, and trace lengths for reasonable data sources. In terms of user-specified nondeterminism, the constant scaling properties for increasing lengths of `do` and `do until` statements demonstrates an ability to handle user-specified nondeterminism without incurring significant overhead. This is because the querying system translates the `do` statement into `do until *` for some uninterpreted Boolean denoted `*` that can take any value at each timestep. If the user selects this option, a sequence of `do` statements will explore all possible sequences that result in a potential match to the labeled trace, instead of specifying values of Boolean variables determining the control flow of the scenario at each time step.

Table 2.1: Scalability Test Results for (N) DoUntil Statements (in sec)

| DoUntil (N) | 10 Timesteps | 20 Timesteps | 30 Timesteps | 40 Timesteps |
|---|---|---|---|---|
| 1 | $1.96 \pm 0.07$ | $2.31 \pm 0.05$ | $2.69 \pm 0.09$ | $3.35 \pm 0.09$ |
| 2 | $2.04 \pm 0.06$ | $2.40 \pm 0.08$ | $3.02 \pm 0.04$ | $3.60 \pm 0.12$ |
| 3 | $2.19 \pm 0.06$ | $2.62 \pm 0.10$ | $3.25 \pm 0.02$ | $3.92 \pm 0.06$ |
| 4 | $2.28 \pm 0.05$ | $2.76 \pm 0.10$ | $3.45 \pm 0.07$ | $4.27 \pm 0.07$ |

Table 2.2: Scalability Test Results for (N) Do Statements (in sec)

| Do (N) | 10 Timesteps | 20 Timesteps | 30 Timesteps | 40 Timesteps |
|---|---|---|---|---|
| 1 | $2.15 \pm 0.05$ | $2.74 \pm 0.10$ | $3.56 \pm 0.06$ | $4.92 \pm 0.10$ |
| 2 | $2.10 \pm 0.06$ | $2.61 \pm 0.10$ | $3.56 \pm 0.13$ | $5.25 \pm 0.18$ |
| 3 | $2.16 \pm 0.05$ | $2.75 \pm 0.09$ | $3.87 \pm 0.12$ | $5.73 \pm 0.18$ |
| 4 | $2.23 \pm 0.06$ | $2.94 \pm 0.10$ | $4.10 \pm 0.17$ | $4.27 \pm 0.07$ |

Table 2.3: Scalability Test Results for (N)ested Try-Interrupt (in sec)

| Depth (N) | 10 Timesteps | 20 Timesteps | 30 Timesteps | 40 Timesteps |
|---|---|---|---|---|
| 1 | $2.91 \pm 0.06$ | $8.39 \pm 0.36$ | $26.72 \pm 0.81$ | – |
| 2 | $6.33 \pm 0.25$ | $35.07 \pm 2.50$ | – | – |
| 3 | $11.56 \pm 0.16$ | – | – | – |
| 4 | $21.95 \pm 0.36$ | – | – | – |

Table 2.4: Scalability Test Results for Try-(N)terrupt (in sec)

| Handlers (N) | 10 Timesteps | 20 Timesteps | 30 Timesteps | 40 Timesteps |
|---|---|---|---|---|
| 1 | $3.31 \pm 0.22$ | $9.80 \pm 0.63$ | $27.67 \pm 1.99$ | – |
| 2 | $5.23 \pm 0.58$ | $29.24 \pm 1.27$ | – | – |
| 3 | $10.29 \pm 0.30$ | – | – | – |
| 4 | $15.31 \pm 1.78$ | – | – | – |

Meanwhile, the evaluations demonstrate support for the entire fragment of SCENIC. The support of longer sequences `do` and `do until` statements with constant overhead extends efficient system support to a sizable amount of behavior definitions. In terms of nested try-interrupt and multiple interrupt handler cases, we observe naturally occurring bounds on the number of useful interrupt conditions in a naturalistic driving behavior representation. More concretely, several large-scale autonomous driving challenge datasets involving scenario programs [82] leverage 1-2 interrupt handlers and 0-1 nested interrupts at most. Let $c$ denote a small constant relative to $k$, $m$, and $T$. For $N=c$ interrupt handlers or levels of nesting and common ranges of $T$ we previously identified, from a simple linear regression over our evaluations we observe maximum runtime can be approximated as a linear function of $T$. This implies that our `nested try-interrupt` and `multiple interrupt handler` statements can be processed in an approximately linear manner with respect to the input trace length $T$. More precisely, suppose $e_T$ denotes the worst-case runtime of processing a trace of length $T$. Based on a simple geometric series modeling of trace length $T$ against evaluation time $e_T$, the runtime for checking each trace as one more timestep is added appears to scale according to the rate $e_{T+1} \approx 1.09 * e_T$.

In cases with significantly longer traces as a result of higher polling rates, by the earlier discussion we propose that the labeled trace behavior classification can continue to operate on a lower sampling rate in Hz. More importantly, many objects in autonomous vehicles traces only remain in range of the data collection vehicle for relatively short amounts of time in comparison to the full scene length. Depending on the sensor data domain and the underlying distribution of individual trace lengths, additional costs associated with longer traces can potentially be amortized to provide stronger efficiency guarantees. *As a result, we propose the entire* SCENIC *fragment can be efficiently processed even when modeling an exponentially growing collection of traces or nondeterminstic outcomes over the state space.*

We observe that the evaluations presented require the usage of a worst-case implementation to provide more reasonable and predictable comparisons across values of $N$ and $T$ for different statements. In order to guarantee the entire trace is processed, we propose the notion of a maximum contiguous match threshold. Within this definition of a match, for some user-specified threshold $m$, the length of a match must be larger than $m$. It is worth noting that the accuracy and querying guarantees of the algorithm depend on the accuracy of the labels and behavior classification system.

In the current state, two key limitations of the algorithm include the limitations of the current SCENIC fragment and the scalability of the current UCLID5-based implementation. The fragment of SCENIC for which the algorithm is currently supported excludes certain conditional and iterative structures, due to the complexity of efficiently representing these structures within a hierarchical UCLID5 state machine. Writing *for* and *while* loops that precisely match labeled traces proved to be an extremely difficult task for SCENIC experts without any information about the sensor data. We discuss several opportunities to augment the system to assist users with the process of writing queries within the Future Work (Section 3.2). With regards to system scalability, the current system generates a new UCLID5 program each time a trace $l$ is checked against a program model $P$. The runtime charts included in Figure 2.5 demonstrate the challenges of scaling this system to terabytes without access to distributed computation, as each of the runtime curves converges to a clear nonzero runtime $e_{min} \approx$ 2.0 sec. In the Future Work, we will propose several optimizations to improve the efficiency of the match searching and UCLID5 encoding processes. These proposals aim to improve the system usability and ensure the worst-case runtimes are seldom reached for individual correspondences.

# Chapter 3

# Future Work and Conclusion

## 3.1 Human Subject Study Proposal

**Setup** In the same manner as Experiment I, we define 5 scenarios of interest but instead ask 3 human participants to query matching time series data points for each of the 5 by hand. In order to acquire the most accurate subsets, we define the test set as the intersection of each human subject's hand-queried test set. To minimize errors due to time constraints, the participant is then given an opportunity to confirm or overturn each response in an unlabeled set of responses. This set of responses corresponds to the set for which their classification differed from that of the querying algorithm. At this point, we compare these subsets to the subsets returned by the algorithm.

**Scenarios** We define five scenarios in a range of realistic traffic situations at differing risk levels and frequencies within the dataset. We ensure that each of the five scenarios defined exists in the labeled dataset nuScenes in some capacity, and we provide natural language descriptions and corresponding Scenic encodings for three of the five scenarios listed below.

(1) Jaywalking pedestrian triggered sudden braking

(2) Yielded to another vehicle while making right turn

(3) Activated braking in response to braking leading vehicle

**Data** In order to allow human subjects to query time series data points, we provide a selection of RGB videos captured from the driver's view (front camera) of the vehicle, based on the corresponding camera angles accessible from the nuScenes dataset. The map information and traffic flow information provided from the nuScenes dataset allow the Scenic scenarios to use information involving map information and traffic flow directions. Object classes including vehicles, pedestrians, and static objects are captured by the data collecting vehicle, referred to as the 'ego' vehicle in the labeled data and querying scenario program. After filtering certain classes of scenes that are less relevant to the scenario queries (i.e. parking lot navigation), we provide the users with a randomly selected subset of videos. We believe this dataset is of reasonable size for humans to manually query from the dataset.

## 3.2 Future Work

Future work may involve creating a more general interface for the algorithm to directly integrate with larger-scale datasets like the Google Open-X Embodiment dataset [35]. For specific domains such as autonomous vehicles, a base set of interesting driving scenarios implemented in SCENIC could be used to provide measures of test coverage and missing scenarios in a real-world AV sensor dataset [82]. To make the system easier to set up and use across different domains, it may be interesting to explore extensions of this work that could apply to unlabeled time series data or collections of multiple datasets. For instance, implementing unsupervised or self-supervised labeling approaches would allow the system to immediately operate and deliver meaningful results in domains with more constrained access to data labeling and open-source data classification systems. The generalizability of the algorithm and the SCENIC language presents exciting opportunities to explore applications of this work in new domains.

In terms of system usability, one promising avenue involves integrating scenario generation systems that generate SCENIC code from natural language [45]. This presents the opportunity to perform end-to-end time series retrieval of real-world sensor data based on natural language descriptions of inputs, abstracting away the SCENIC development process for less formal use cases. Natural language sim-to-real querying would present a promising avenue for individuals from less technical backgrounds to more robustly validate autonomous systems and sensor data they collect, while improving transparency into the contents of large-scale datasets and transferability of simulation testing results to information about real-world performance. Several quick implementation changes, including configuring UCLID5 programs to accept multiple traces as inputs, may provide significant runtime boosts over the existing system to bypass the initial startup costs associated with initializing UCLID5 modules. In addition, the object correspondence problem can be divided into multiple independent subproblems based on independent object classes or behavior definitions. For scenario programs with larger numbers of agents, this reduce the number of candidate correspondences to check by several orders of magnitude. As the evaluations included in Experiment II focus on worst-case runtimes enforced over synthetic predicate streams, the runtimes are not representative of average bounded model checking evaluations of a match between $P$ and $l$. A more precise measure of scalability would involve large-scale evaluations on full real-world sensor datasets with distributed systems.

A far more critical line of future work involves assisting users with the process of constructing and modifying queries to retrieve larger distributions of results. As a result of the strict distributions SCENIC programs define, even slight changes to random variables or conditional values can impact the retrieved results of the querying process. Implementing solver or learning-based tools to suggest program and program parameter adjustments to improve querying results will improve the usability of the algorithm across a wide range of domains. In addition, more precise UCLID5 assertions defining the boundary between a match and a failure may improve. For instance, reweighting the contribution of certain labels towards the definition of a match based on the frequency may assist with the discovery of rarer scenarios (i.e. lane change scenarios) that may not appear as frequently in the dataset of labeled traces.

## 3.3 Conclusion

In this work, we proposed an algorithm using a scenario program encoded in SCENIC to query over a labeled time series dataset. This algorithm can be used to close the gap between simulation and real-world testing by identifying real-world data points in a time series dataset corresponding to a dynamic, multi-agent evaluation scenario for a cyber-physical system. More broadly, the algorithm formalizes an efficient process of exploring and understanding the contents of a real-world dataset. This algorithm supports dynamic scenarios capable of querying for the nondeterminism the SCENIC language allows users to express. In addition, the generalizability of the approach allows it to be adapted to a variety of domains, including autonomous vehicles and indoor robotics. With the scalability, transparency, and robustness of this algorithm, we hope to see an accelerated pace of development and understanding of more comprehensive datasets and learning-based tasks within cyber-physical systems.

BIBLIOGRAPHY 43[77] S. Gao, S. Kong, and E. M. Clarke, "Dreal: An smt solver for nonlinear theories over the reals," in *International conference on automated deduction*, Springer, 2013, pp. 208–214.

[78] N. Bjørner, L. de Moura, L. Nachmanson, and C. M. Wintersteiger, "Programming z3," *Engineering Trustworthy Software Systems: 4th International School, SETSS 2018, Chongqing, China, April 7–12, 2018, Tutorial Lectures 4*, pp. 148–201, 2019.

[79] M.-F. Chang, J. Lambert, P. Sangkloy, *et al.*, "Argoverse: 3d tracking and forecasting with rich maps," in *Proceedings of the IEEE/CVF conference on computer vision and pattern recognition*, 2019, pp. 8748–8757.

[80] B. Wilson, W. Qi, T. Agarwal, *et al.*, "Argoverse 2: Next generation datasets for self-driving perception and forecasting," in *Thirty-fifth Conference on Neural Information Processing Systems Datasets and Benchmarks Track (Round 2)*, 2021.

[81] H. Caesar, V. Bankiti, A. H. Lang, *et al.*, "Nuscenes: A multimodal dataset for autonomous driving," in *Proceedings of the IEEE/CVF conference on computer vision and pattern recognition*, 2020, pp. 11 621–11 631.

[82] B. Osiński, P. Miłoś, A. Jakubowski, *et al.*, "Carla real traffic scenarios–novel training ground and benchmark for autonomous driving," *arXiv preprint arXiv:2012.11329*, 2020.

[83] A. Cremers and S. Ginsburg, "Context-free grammar forms," *Journal of Computer and System Sciences*, vol. 11, no. 1, pp. 86–117, 1975.

[84] D. D. McCracken and E. D. Reilly, "Backus-naur form (bnf)," in *Encyclopedia of Computer Science*, 2003, pp. 129–131.

[85] W. Wieczorek, O. Unold, and Ł. Strak, "Parsing expression grammars and their induction algorithm," *Applied Sciences*, vol. 10, no. 23, p. 8747, 2020.

[86] D. Harel, "Statecharts: A visual formalism for complex systems," *Science of computer programming*, vol. 8, no. 3, pp. 231–274, 1987.

[87] A. Roques and P. Contributors, *PlantUML Software*. [Online]. Available: https://github.com/plantuml/plantuml.

[88] A. Decan and T. Mens, "Sismic—a python library for statechart execution and testing," *SoftwareX*, vol. 12, p. 100 590, 2020.

# Appendix A

# Additional Materials

## A.1 Scenic Fragment Extended BNF Grammar

**Formalizing Scenic Fragment as Grammar**

Formalizing the fragment of the SCENIC language for which labeled trace querying is supported is integral to ensuring syntax is properly parsed into the interrupt-driven hierarchical finite state machine (IHEFSM) definition. Context-free grammars [83] such as the Backus-Naur Form (BNF) [84] or Parsing Expression Grammar (PEG) [85] provide a structured and systematic way to describe the syntax of programming languages, including SCENIC. By defining a grammar for a fragment of SCENIC considered for time series data retrieval, it becomes possible to parse and analyze SCENIC code programmatically. This streamlines programmatic tasks including code generation, analysis, and transformation. This formalization enables tools and systems to interact with SCENIC code more reliably and efficiently, laying the groundwork for automated processing and enhanced language features.

**Context-Free Grammars in Programming Language Formalization**

Backus-Naur Form (BNF) and its variants, such as Extended Backus-Naur Form (EBNF) [68], are foundational in the formalization of programming languages. BNF, introduced in the 1960s, offers a concise notation for defining the syntax of programming languages [84]. This allows for ease of understanding, implementing, and communicating language specifications. EBNF enhances BNF Grammar by introducing additional syntax constructs (including optional elements, repetition, flattened lists, and grouping) that simplify grammar definitions and make them more readable and usable. Compared to BNF, EBNF includes reduced ambiguity and a more compact representation of complex syntactic patterns. These features make EBNF preferable to BNF for this specific use case. Since the full SCENIC language and the Python release it was built on both partially employ EBNF-based CFGs, the EBNF seems to be a reasonable and readable starting point for parsing IHEFSM from a context-free grammar of the SCENIC fragment.[1]

---

[1] The Python3 grammar is defined in a combination of EBNF and PEG, while SCENIC is defined in PEG. We define a EBNF Grammar to represent our fragment of SCENIC for this algorithm.

## Design Considerations of Scenic EBNF Grammar

The design of the EBNF grammar for a fragment of the SCENIC language is driven by a need to capture the complex constructs needed to define hierarchical IHEFSMs. Key considerations included the flexibility to define interesting behaviors, the ability to express conditional logic, and the mechanisms for describing state transitions and actions. The grammar supports various statements, such as do, take, try-interrupt, conditional, abort, and terminate statements, each catering to different aspects of HSM definition. Do statements allow for the specification of sustained behaviors, take statements describe immediate actions, and try-interrupt statements enable preemptive behavior switching based on Boolean conditions. The inclusion of abort and terminate statements support more complex and realistic control flow mechanisms.

## Parsing EBNF Grammar into IHEFSM

The algorithm to parse the EBNF grammar of the SCENIC language fragment is designed to systematically interpret SCENIC code and construct a corresponding hierarchical FSM. It operates in three stages: (1) lexical analysis tokenizes the input code, (2) syntactic analysis parses these tokens based on the EBNF grammar to build an abstract syntax tree (AST), (3) semantic analysis traverses the AST in a sequential recursive fashion to construct the IHEFSM. The recursive nature of the algorithm facilitates the automatic generation of IHEFSMs for all programs expressible within the predefined fragment of SCENIC, streamlining the process of constructing a bounded model checking problem for UCLID5.

## EBNF Grammar for Scenic Fragment

```
finite-state-machine ::= { behavior-definition | statement-sequence };

behavior-definition ::= "behavior" identifier "{" statement-sequence "}";

statement-sequence ::= statement { "next" statement } ;

statement ::= do-statement
            | take-statement
            | try-interrupt-statement
            | conditional-statement
            | terminate-statement
            | choose-statement
            | shuffle-statement
            | assignment-statement;

do-statement ::= "do" identifier ["until" condition] ["do" action];

take-statement ::= "take" action;

try-interrupt-statement ::= "try:" statement-sequence { "interrupt
```

```
                              when" condition statement-sequence };

conditional-statement ::= "if" condition ":" statement-sequence
                          { "elif" condition ":" statement-sequence }
                          [ "else:" statement-sequence ];

abort-statement ::= "abort";

terminate-statement ::= "terminate";

choose-statement ::= "choose" identifiers

shuffle-statement ::= "shuffle" identifiers

assignment-statement ::= identifier {"," identifier} "=" value {"," value};

condition ::= boolean-expression;

boolean-expression ::= identifier
                     | boolean-literal
                     | boolean-expression logical-operator boolean-expression
                     | "not" boolean-expression;

logical-operator ::= "and" | "or";

boolean-literal ::= "True" | "False";

action ::= atomic-behavior | identifier | action-sequence;

atomic-behavior ::= "FollowLaneBehavior"
                  | "TurnLeftBehavior"
                  | "TurnRightBehavior"
                  | "BrakingBehavior"
                  | "AccelerateForwardBehavior"
                  | "LaneChangeBehavior";

action-sequence ::= action { "," action };

# lower-level constructs below

identifier ::= letter | "_" { letter | digit | "_" };

identifiers ::= identifier {"," identifier}
```

```
value ::= identifier | number | string | boolean-literal;

number ::= digit { digit };

string ::= "\"" { letter | digit | "_" | " " } "\"";

letter ::= "a" ... "z" | "A" ... "Z";

digit ::= "0" ... "9";
```

## A.2 Scenic to UCLID5 Translation

The following algorithms assist with the generation of interrupt-driven, hierarchical extended finite state machine models of SCENIC programs in UCLID5 code.

---
**Algorithm 6** ParseName (SCENIC to UCLID5 Helper)
---
**Input**: abstract syntax tree of behavior definition (*behavior*)
**Output**: symbolic IHEFSM representation and UCLID5 hierarchical program encoding for the behavior definition body input
 1: $seq \leftarrow GetBehaviorDefinitionBody(behavior)$ // body is *list* of *AST* statements
 2: $program \leftarrow NewUclidProgram()$
 3: $child\_modules, program \leftarrow ParseSequence(seq, program)$ // parent tracks active child
 4: $program \leftarrow NewUclidModule(behavior\_name, child\_modules, program)$ // add module
 5: **return** $program$ // IHEFSM code for all modules of UCLID5 program
---

---
**Algorithm 7** ParseSequence (SCENIC to UCLID5 Helper)
---
**Input**: sequence of SCENIC *AST* statements (*sequence*), UCLID5 code (*program*)
**Output**: symbolic IHEFSM representation of the SCENIC program and UCLID5 hierarchical program encoding the representation
 1: $child\_modules \leftarrow [\,]$ // add module for each statement in sequence
 2: **for** *stmt* in *sequence* **do**
 3:     $module, program \leftarrow ParseStatement(stmt, program)$
 4:     $child\_modules.append(module)$
 5: $program \leftarrow NewUclidModule(stmt, child\_modules, program)$
 6: **return** $child\_modules, program$ // each statement in sequence has corresponding module
---

Algorithm 6 (*ParseName*) creates the top-level UCLID5 module representing the state that contains all trace-generating logic until the program terminates. Algorithm 7 (*ParseSequence*) and Algorithm 8 (*ParseStatement*) recursively call each other to parse out sequences of SCENIC

---

**Algorithm 8** ParseStatement (Scenic to UCLID5 Helper)

---

**Input**: Single Scenic *AST* statement (*statement*), UCLID5 code (*program*)
**Output**: Symbolic IHEFSM representation of the Scenic program and UCLID5 hierarchical program encoding the representation

1: **if** $IsAtomic(statement)$ **then**
2:     **return** $\varnothing, NewUclidModule(statement, \varnothing, program)$
3: $module, program \leftarrow ParseSequence(statement.body, program)$ // body is list of ASTs
4: $program \leftarrow NewUclidModule(statement, module, program)$
5: // $NewUclidModule$ handles logic for DoUntil, TryInterrupt, Try, Interrupt, etc.
6: **return** $[child\_module], program$ // statement has container module

---

statements until reaching any base cases (`do` and `do until` statements with behavior outputs contained in the set of possible output trace labels).

## A.3 Experiment I Example Query

**Scenic Query Program**

```
behavior EgoBehavior():
    try:
        do FollowLaneBehavior()
    interrupt when (distance from self to ped) < Range(1,10):
        do BrakeBehavior()

ego = new Car with behavior EgoBehavior()
ped = new Pedestrian
```

**Predicate Abstraction of Scenic Program**

The predicate abstraction process automatically extracts the set of Boolean variable predicates directing the flow of execution for the SCENIC program. For each extracted predicate such as `nusc_cond_interrupt_1_1`, it generates a stream of Boolean outputs for each timestep based on the real-world labeled trace data.

```
behavior EgoBehavior():
    try:
        do FollowLaneBehavior()
    interrupt when (nusc_cond_interrupt_1_1):
        do BrakeBehavior()

ego = new Car with behavior EgoBehavior()
ped = new Pedestrian
```

## Generated Statecharts Representation of Scenic Program

Figure A.1 is an automatically generated IHEFSM representation of the SCENIC program in statecharts [86]. Note that this is a byproduct of the translation process constructed for interpretability, but the translation from SCENIC to UCLID5 is direct. The rendering is created with a PlantUML [87] wrapper created through the Sismic [88] statecharts library.

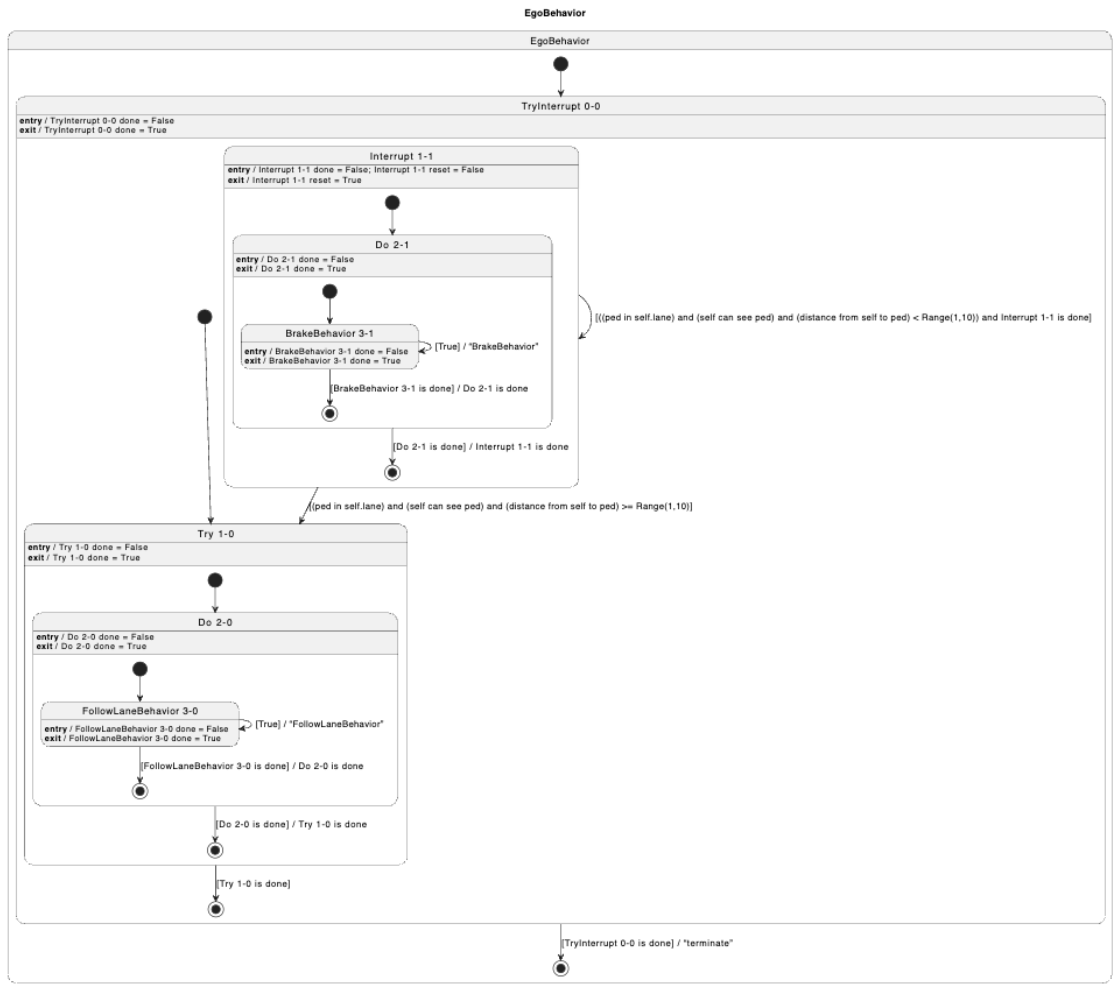

Figure A.1: The Statecharts representation of the SCENIC program contains hierarchical states `EgoBehavior`, `TryInterrupt`, `Try`, `Interrupt`, `Do`, `FollowLaneBehavior`, `BrakeBehavior` corresponding to hierarchical structure of the SCENIC program. Note that within this representation, actions at each timestep are output by transitions at the deepest states of the interrupt-driven hierarchical state machine.

## Generated UCLID5 Hierarchical Encoding of Scenic Program

```
module FollowLane_3_0 {
    type atomic_t = input_generator.atomic_t;
    type data_t = input_generator.data_t;
    type status_t = input_generator.status_t;
    type reset_t = input_generator.reset_t;

    input nusc_cond_interrupt_1_1 : data_t;
    input nusc_cond_do_2_0 : data_t;
    input nusc_cond_do_2_1 : data_t;
    sharedvar hfsm_trace : atomic_t;
    sharedvar reset_target : reset_t;

    output status_followlane_3_0 : status_t;

    procedure reset_3_0()
        modifies status_followlane_3_0;
    {
        status_followlane_3_0 = start;
    }

    init {
        status_followlane_3_0 = start;
    }
    next {
        hfsm_trace' = FollowLaneBehavior;
        status_followlane_3_0' = progress;
    }
}

module Brake_3_1 {
    type atomic_t = input_generator.atomic_t;
    type data_t = input_generator.data_t;
    type status_t = input_generator.status_t;
    type reset_t = input_generator.reset_t;

    input nusc_cond_interrupt_1_1 : data_t;
    input nusc_cond_do_2_0 : data_t;
    input nusc_cond_do_2_1 : data_t;
    sharedvar hfsm_trace : atomic_t;
    sharedvar reset_target : reset_t;

    output status_brake_3_1 : status_t;
```

```
    procedure reset_3_1()
        modifies status_brake_3_1;
    {
        status_brake_3_1 = start;
    }

    init {
        status_brake_3_1 = start;
    }
    next {
        hfsm_trace' = BrakeBehavior;
        status_brake_3_1' = progress;
    }
}

module Do_2_0 {
    type atomic_t = input_generator.atomic_t;
    type data_t = input_generator.data_t;
    type status_t = input_generator.status_t;
    type reset_t = input_generator.reset_t;

    input nusc_cond_interrupt_1_1 : data_t;
    input nusc_cond_do_2_0 : data_t;
    input nusc_cond_do_2_1 : data_t;
    sharedvar hfsm_trace : atomic_t;
    sharedvar reset_target : reset_t;

    var status_followlane_3_0 : status_t;
    output status_do_2_0 : status_t;

    instance followlane_3_0: FollowLane_3_0(
        hfsm_trace : (hfsm_trace),
        reset_target : (reset_target),
        nusc_cond_interrupt_1_1 : (nusc_cond_interrupt_1_1),
        nusc_cond_do_2_0 : (nusc_cond_do_2_0),
        nusc_cond_do_2_1 : (nusc_cond_do_2_1),
        status_followlane_3_0 : (status_followlane_3_0)
    );

    procedure reset_2_0()
        modifies status_do_2_0, followlane_3_0;
    {
        status_do_2_0 = start;
        call followlane_3_0.reset_3_0();
```

```
        }

        init {
            status_do_2_0 = start;
        }
        next {
            case
                (status_do_2_0 != end) : {
                    next(followlane_3_0);

                    if (nusc_cond_do_2_0) {
                        status_do_2_0' = end;
                    } else {
                        status_do_2_0' = progress;
                    }
                }
            esac
        }
    }

    module Do_2_1 {
        type atomic_t = input_generator.atomic_t;
        type data_t = input_generator.data_t;
        type status_t = input_generator.status_t;
        type reset_t = input_generator.reset_t;

        input nusc_cond_interrupt_1_1 : data_t;
        input nusc_cond_do_2_0 : data_t;
        input nusc_cond_do_2_1 : data_t;
        sharedvar hfsm_trace : atomic_t;
        sharedvar reset_target : reset_t;

        var status_brake_3_1 : status_t;
        output status_do_2_1 : status_t;

        instance brake_3_1: Brake_3_1(
            hfsm_trace : (hfsm_trace),
            reset_target : (reset_target),
            nusc_cond_interrupt_1_1 : (nusc_cond_interrupt_1_1),
            nusc_cond_do_2_0 : (nusc_cond_do_2_0),
            nusc_cond_do_2_1 : (nusc_cond_do_2_1),
            status_brake_3_1 : (status_brake_3_1)
        );
```

```
    procedure reset_2_1()
        modifies status_do_2_1, brake_3_1;
    {
        status_do_2_1 = start;
        call brake_3_1.reset_3_1();
    }

    init {
        status_do_2_1 = start;
    }
    next {
        case
            (status_do_2_1 != end) : {
                next(brake_3_1);

                if (nusc_cond_do_2_1) {
                    status_do_2_1' = end;
                } else {
                    status_do_2_1' = progress;
                }
            }
        esac
    }
}

module Try_1_0 {
    type atomic_t = input_generator.atomic_t;
    type data_t = input_generator.data_t;
    type status_t = input_generator.status_t;
    type reset_t = input_generator.reset_t;

    input nusc_cond_interrupt_1_1 : data_t;
    input nusc_cond_do_2_0 : data_t;
    input nusc_cond_do_2_1 : data_t;
    sharedvar hfsm_trace : atomic_t;
    sharedvar reset_target : reset_t;

    var status_do_2_0 : status_t;
    output status_try_1_0 : status_t;

    instance do_2_0: Do_2_0(
        hfsm_trace : (hfsm_trace),
        reset_target : (reset_target),
        nusc_cond_interrupt_1_1 : (nusc_cond_interrupt_1_1),
```

```
        nusc_cond_do_2_0 : (nusc_cond_do_2_0),
        nusc_cond_do_2_1 : (nusc_cond_do_2_1),
        status_do_2_0 : (status_do_2_0)
    );

    procedure reset_1_0()
        modifies status_try_1_0, do_2_0;
    {
        status_try_1_0 = start;
        call do_2_0.reset_2_0();
    }

    init {
        status_try_1_0 = start;
    }
    next {
        case
            (status_try_1_0 != end) : {
                next(do_2_0);

                if (status_do_2_0' == end) {
                    status_try_1_0' = end;
                } else {
                    status_try_1_0' = progress;
                }
            }
        esac
    }
}

module Interrupt_1_1 {
    type atomic_t = input_generator.atomic_t;
    type data_t = input_generator.data_t;
    type status_t = input_generator.status_t;
    type reset_t = input_generator.reset_t;

    input nusc_cond_interrupt_1_1 : data_t;
    input nusc_cond_do_2_0 : data_t;
    input nusc_cond_do_2_1 : data_t;
    sharedvar hfsm_trace : atomic_t;
    sharedvar reset_target : reset_t;

    var status_do_2_1 : status_t;
    output status_interrupt_1_1 : status_t;
```

```
    instance do_2_1: Do_2_1(
        hfsm_trace : (hfsm_trace),
        reset_target : (reset_target),
        nusc_cond_interrupt_1_1 : (nusc_cond_interrupt_1_1),
        nusc_cond_do_2_0 : (nusc_cond_do_2_0),
        nusc_cond_do_2_1 : (nusc_cond_do_2_1),
        status_do_2_1 : (status_do_2_1)
    );

    procedure reset_1_1()
        modifies status_interrupt_1_1, do_2_1;
    {
        status_interrupt_1_1 = start;
        call do_2_1.reset_2_1();
    }

    init {
        status_interrupt_1_1 = start;
    }
    next {
        case
            (status_interrupt_1_1 != end) : {
                next(do_2_1);

                if (status_do_2_1' == end) {
                    status_interrupt_1_1' = end;
                } else {
                    status_interrupt_1_1' = progress;
                }
            }
        esac
    }
}

module TryInterrupt_0_0 {
    type atomic_t = input_generator.atomic_t;
    type data_t = input_generator.data_t;
    type status_t = input_generator.status_t;
    type reset_t = input_generator.reset_t;
    type state_t = enum {TRY_1_0, INT_1_1};

    input nusc_cond_interrupt_1_1 : data_t;
    input nusc_cond_do_2_0 : data_t;
```

```
input nusc_cond_do_2_1 : data_t;
sharedvar hfsm_trace : atomic_t;
sharedvar reset_target : reset_t;

var status_try_1_0 : status_t;
var status_interrupt_1_1 : status_t;
output status_tryinterrupt_0_0 : status_t;

var state_0_0 : state_t;

instance try_1_0: Try_1_0(
    hfsm_trace : (hfsm_trace),
    reset_target : (reset_target),
    nusc_cond_interrupt_1_1 : (nusc_cond_interrupt_1_1),
    nusc_cond_do_2_0 : (nusc_cond_do_2_0),
    nusc_cond_do_2_1 : (nusc_cond_do_2_1),
    status_try_1_0 : (status_try_1_0)
);
instance interrupt_1_1: Interrupt_1_1(
    hfsm_trace : (hfsm_trace),
    reset_target : (reset_target),
    nusc_cond_interrupt_1_1 : (nusc_cond_interrupt_1_1),
    nusc_cond_do_2_0 : (nusc_cond_do_2_0),
    nusc_cond_do_2_1 : (nusc_cond_do_2_1),
    status_interrupt_1_1 : (status_interrupt_1_1)
);

procedure reset_0_0()
    modifies status_tryinterrupt_0_0, try_1_0, interrupt_1_1;
{
    status_tryinterrupt_0_0 = start;
    call try_1_0.reset_1_0();
    call interrupt_1_1.reset_1_1();
}

init {
    status_tryinterrupt_0_0 = start;
    state_0_0 = TRY_1_0;
}
next {
    case
        (status_tryinterrupt_0_0 != end) : {
            case
            (state_0_0 == INT_1_1) : {
```

```
            next(interrupt_1_1);
            case
                (status_interrupt_1_1' != end) : {
                    state_0_0' = INT_1_1;
                }
                default : {
                    state_0_0' = TRY_1_0;
                }
            esac
            status_tryinterrupt_0_0' = progress;
        }
        (state_0_0 == TRY_1_0) : {
            next(try_1_0);
            if (status_try_1_0' == end) {
                status_tryinterrupt_0_0' = end;
            } else {
                status_tryinterrupt_0_0' = progress;
                case
                    (nusc_cond_interrupt_1_1) : {
                        state_0_0' = INT_1_1;
                    }
                    default : {
                        state_0_0' = TRY_1_0;
                    }
                esac
            }
        }
        esac

        case
            (reset_target == no_reset) : {
                case
                    (state_0_0 == INT_1_1 && state_0_0' == TRY_1_0) : {
                        reset_target' = reset_interrupt_1_1;
                    }
                esac
            }
            (reset_target == reset_interrupt_1_1) : {
                call interrupt_1_1.reset_1_1();
                reset_target' = no_reset;
            }
        esac
    }
esac
```

```
        }
}

module EgoBehavior {
    type atomic_t = input_generator.atomic_t;
    type data_t = input_generator.data_t;
    type status_t = input_generator.status_t;
    type reset_t = input_generator.reset_t;

    input nusc_cond_interrupt_1_1 : data_t;
    input nusc_cond_do_2_0 : data_t;
    input nusc_cond_do_2_1 : data_t;
    sharedvar hfsm_trace : atomic_t;
    sharedvar reset_target : reset_t;

    var status_tryinterrupt_0_0 : status_t;

    instance tryinterrupt_0_0: TryInterrupt_0_0(
        hfsm_trace : (hfsm_trace),
        reset_target : (reset_target),
        nusc_cond_interrupt_1_1 : (nusc_cond_interrupt_1_1),
        nusc_cond_do_2_0 : (nusc_cond_do_2_0),
        nusc_cond_do_2_1 : (nusc_cond_do_2_1),
        status_tryinterrupt_0_0 : (status_tryinterrupt_0_0)
    );
    instance terminate: Terminate(
        hfsm_trace : (hfsm_trace)
    );

    init {
        reset_target = no_reset;
    }
    next {
        case
            (status_tryinterrupt_0_0 != end) : {
                next(tryinterrupt_0_0);
            }
            default : {
                next(terminate);
            }
        esac
    }
}
```

## A.4  Experiment II Evaluation Programs

**Experimental Overview**

The following programs were used in the scalability analysis. Each of the four program formats can be scaled to any number of statements (N), but we limit our experiments to the 1-4 range due to computational constraints.

## Nested Try

**Nested Try 1**

```
behavior TestParseBehavior():
    try:
        do FollowLaneBehavior() until cond
    interrupt when cond:
        do BrakingBehavior() until cond

ego = new Car with behavior TestParseBehavior()
```

**Nested Try 2**

```
behavior TestParseBehavior():
    try:
        try:
            do FollowLaneBehavior() until cond
        interrupt when cond:
            do TurnRightBehavior() until cond
    interrupt when cond:
        do BrakingBehavior() until cond

ego = new Car with behavior TestParseBehavior()
```

**Nested Try 3**

```
behavior TestParseBehavior():
    try:
        try:
            try:
                do FollowLaneBehavior() until cond
            interrupt when cond:
                do TurnLeftBehavior() until cond
        interrupt when cond:
            do TurnRightBehavior() until cond
    interrupt when cond:
```

```
        do BrakingBehavior() until cond

ego = new Car with behavior TestParseBehavior()
```

**Nested Try 4**

```
behavior TestParseBehavior():
    try:
        try:
            try:
                try:
                    do FollowLaneBehavior() until cond
                interrupt when cond:
                    do AccelerateForwardBehavior() until cond
            interrupt when cond:
                do TurnLeftBehavior() until cond
        interrupt when cond:
            do TurnRightBehavior() until cond
    interrupt when cond:
        do BrakingBehavior() until cond

ego = new Car with behavior TestParseBehavior()
```

# Try N

### Try N 1

```
behavior TestParseBehavior():
    try:
        do FollowLaneBehavior() until cond
    interrupt when cond:
        do BrakingBehavior() until cond

ego = new Car with behavior TestParseBehavior()
```

### Try N 2

```
behavior TestParseBehavior():
    try:
        do FollowLaneBehavior() until cond
    interrupt when cond:
        do TurnRightBehavior() until cond
    interrupt when cond:
        do BrakingBehavior() until cond
```

```
ego = new Car with behavior TestParseBehavior()
```

### Try N 3

```
behavior TestParseBehavior():
    try:
        do FollowLaneBehavior() until cond
    interrupt when cond:
        do TurnLeftBehavior() until cond
    interrupt when cond:
        do TurnRightBehavior() until cond
    interrupt when cond:
        do BrakingBehavior() until cond

ego = new Car with behavior TestParseBehavior()
```

### Try N 4

```
behavior TestParseBehavior():
    try:
        do FollowLaneBehavior() until cond
    interrupt when cond:
        do AccelerateForwardBehavior() until cond
    interrupt when cond:
        do TurnLeftBehavior() until cond
    interrupt when cond:
        do TurnRightBehavior() until cond
    interrupt when cond:
        do BrakingBehavior() until cond

ego = new Car with behavior TestParseBehavior()
```

## N DoUntil

### N DoUntil 1

```
behavior TestParseBehavior():
    do FollowLaneBehavior() until cond

ego = new Car with behavior TestParseBehavior()
```

### N DoUntil 2

```
behavior TestParseBehavior():
    do FollowLaneBehavior() until cond
    do BrakingBehavior() until cond

ego = new Car with behavior TestParseBehavior()
```

### N DoUntil 3

```
behavior TestParseBehavior():
    do FollowLaneBehavior() until cond
    do TurnRightBehavior() until cond
    do BrakingBehavior() until cond

ego = new Car with behavior TestParseBehavior()
```

### N DoUntil 4

```
behavior TestParseBehavior():
    do FollowLaneBehavior() until cond
    do TurnLeftBehavior() until cond
    do TurnRightBehavior() until cond
    do BrakingBehavior() until cond

ego = new Car with behavior TestParseBehavior()
```

## N Do

### N Do 1

```
behavior TestParseBehavior():
    do FollowLaneBehavior()

ego = new Car with behavior TestParseBehavior()
```

### N Do 2

```
behavior TestParseBehavior():
    do FollowLaneBehavior()
    do BrakingBehavior()

ego = new Car with behavior TestParseBehavior()
```

**N Do 3**

```
behavior TestParseBehavior():
    do FollowLaneBehavior()
    do TurnRightBehavior()
    do BrakingBehavior()

ego = new Car with behavior TestParseBehavior()
```

**N Do 4**

```
behavior TestParseBehavior():
    do FollowLaneBehavior()
    do TurnLeftBehavior()
    do TurnRightBehavior()
    do BrakingBehavior()

ego = new Car with behavior TestParseBehavior()
```

## A.5 Impracticality of Simulation-Based Approaches

The following algorithm formulation attempts to query from the state representation without leveraging a formal verification system. It fails to scale and is practically tedious to implement, as it can require an exponential number of predicates to be added to an SMT solver and must store an SMT solver per state in object IHEFSM, per object in $l$. Though simulation-based approaches handle simpler cases much more efficiently, many break down once user-specified nondeterminism enters the supported fragment of SCENIC.

After exploring several algorithmic approaches of this nature, we propose that the bounded model checking formulation of the querying problem more reasonably scales to handle cases of nondeterminism in more complex scenarios. Statechart simulation and direct evaluation approaches may be required to generate incredible amounts of possible traces in order to determine if a program $P$ could possibly generate a trace $l$.

**Algorithm 9** Determining if SCENIC program $P$ matches a time series label $l$

**Input** : A SCENIC program $P$, a library of pre-defined behaviors $B$, a time series label $l$
**Output**: Does $l$ match $P$? (True/False)

1: $l^* \leftarrow AugmentLabel(l, B)$ // add behavior trace to the label
2: $AST \leftarrow Compile(P)$ // get abstract syntax tree (AST)
3: $IHFSM \leftarrow ExtractFSM(AST, B)$ // get Interrupt-driven Hierarchical FSM
4: $\phi \leftarrow InitializeSMTFormula(s)$ // To search for a feasible object corresp.
5: **while** $SMTSolver(\phi)$ has a solution **do**
6:    $c \leftarrow SMTSolver(\phi)$ // a feasible object correspondence $c$
7:    $matchFailed \leftarrow$ false
8:    $d_{states} \leftarrow$ {key: agent, value: a set of feasible current states in $IHFSM$}
9:    $d_{constraints} \leftarrow$ {key: state, value: dictionary {key: random variable, value: $\emptyset$}}
10:    **for each** timestep $t$ in the label $l$ **do**
11:      **for each** agent $a$ in the SCENIC program $P$ **do**
12:        $CS \leftarrow d_{states}[a]$ // current states of $a$ in the $IHFSM$
13:        $b_{a,t} \leftarrow$ a set of behaviors taken by $a$ at timestep $t$ in label $l$ according to $c$
14:        $TC_{a,t} \leftarrow$ a set of transition conditions which output $b_{a,t}$ from $CS$
15:        $feasibleNextStates \leftarrow \emptyset$
16:        $nextDict \leftarrow \emptyset$
17:        **for each** transition condition $tc$ from $TC_{a,t}$ **do**
18:           $\psi_{tc} \leftarrow$ SMT encoding of $tc$ with assignments of $a$'s feature values in $l$ at $t$
19:           $rv \leftarrow$ a set of all random variables invoked in $tc$
20:           $\psi \leftarrow \psi_{tc} \wedge$ (SMT formula in $d_{constraints}[s][r]. \forall s \in CS, \forall r \in rv$)
21:           **if** $SMTSolver(\psi)$ is true **then**
22:              $currentState, nextState \leftarrow$ current and next states of $tc$
23:              Add $nextState$ to $feasibleNextStates$
24:              $updatedConstraint \leftarrow (d_{constraints}[currentState][r] \wedge \psi_{tc})$
25:              **if** $nextState \notin nextDict.keys$ **then**
26:                 $\forall r \in rv. \; nextDict[nextState][r] \leftarrow \psi_{tc} \wedge d_{constraints}[currentState][r]$
27:              **else**
28:                 $\forall r \in rv. \; nextDict[nextState][r] \leftarrow nextDict[nextState][r] \vee (\psi_{tc} \wedge d_{constraints}[currentState][r])$
29:        **if** $feasibleNextStates \neq \emptyset$ **then**
30:           $d_{states}[a] \leftarrow feasibleNextStates$
31:           Delete $d_{constraints}.keys \in d_{states}[a]$
32:           Add $nextDict$ to $d_{constraints}[a]$
33:        **else**
34:           $\phi \leftarrow \phi \wedge$ ((partial) correspondence for the *agents* invalidated so far in $c$)
35:           $matchFailed \leftarrow$ true
36:           **break** out of all for loops
37:    **if not** $matchFailed$ **then**
38:      **return** *True*
39: **return** *False*

## A.6 Hierarchical Representation of Scenic Programs

### Translation Overview

Through the following simple SCENIC to statecharts translations, we aim to practially demonstrate the manner in which the hierarchical state machine (IHEFSM) representation of a SCENIC program is hierarchically constructed. All statecharts visualizations continue to be automatically translated.

### Do Statement

**Scenic Code**

```
behavior EgoBehavior():
    do FollowLaneBehavior()

ego = new Car with behavior EgoBehavior()
```

**Statechart Representation**

Note that the `do AtomicBehavior` statement has a a parent `do` state and a child `AtomicBehavior` state. This allows the same representation of `do` statements to be used for atomic behaviors and user-defined behaviors, which may insert a hierarchical state with deeper substates in the place of a single state.

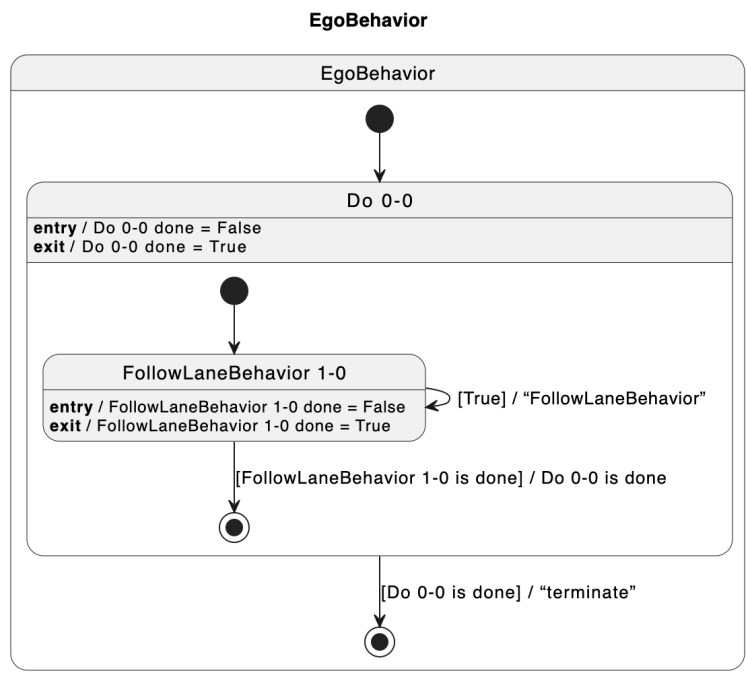

Figure A.2: Statechart representation for a `do` statement.

## Do Until Statement

### Scenic Code

```
behavior EgoBehavior():
    do FollowLaneBehavior() until (distance from self to ped) < Range(1,10)

ego = new Car with behavior EgoBehavior()
ped = new Pedestrian
```

### Statechart Representation

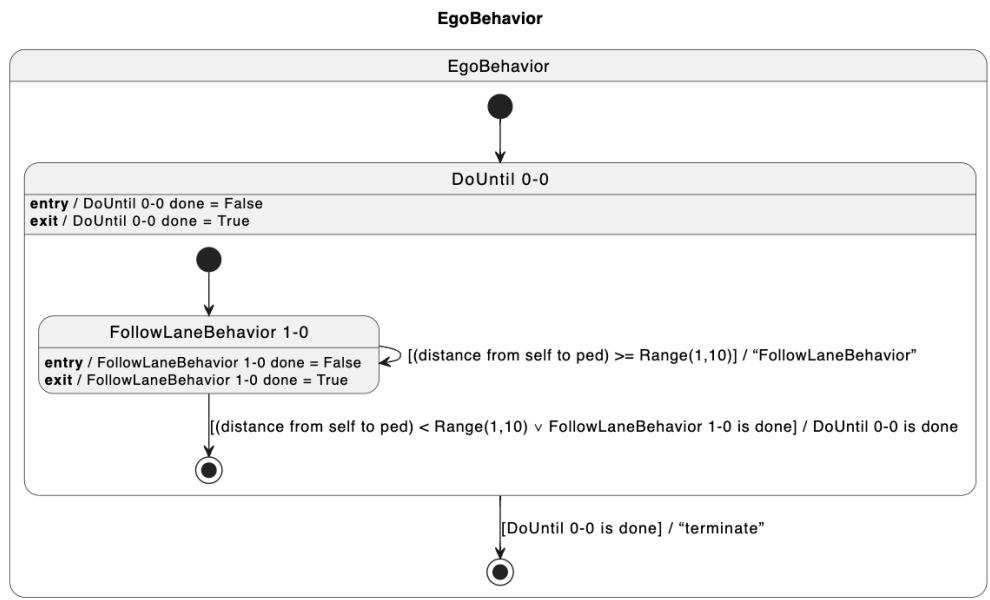

Figure A.3: Statechart representation for a `do until` statement. Observe that only the transition conditions have changed between the `do until` and `do` representations. This allows for ease of introduction of undeclared Boolean variables into UCLID5 programs to allow for the exploration of nondeterministic or undefined termination conditions.

## Sequential Do Statements

**Scenic Code**

```
behavior EgoBehavior():
    do FollowLaneBehavior() until (distance from self to ped) < Range(1,10)
    do BrakingBehavior()

ego = new Car with behavior EgoBehavior()
ped = new Pedestrian
```

**Statechart Representation**

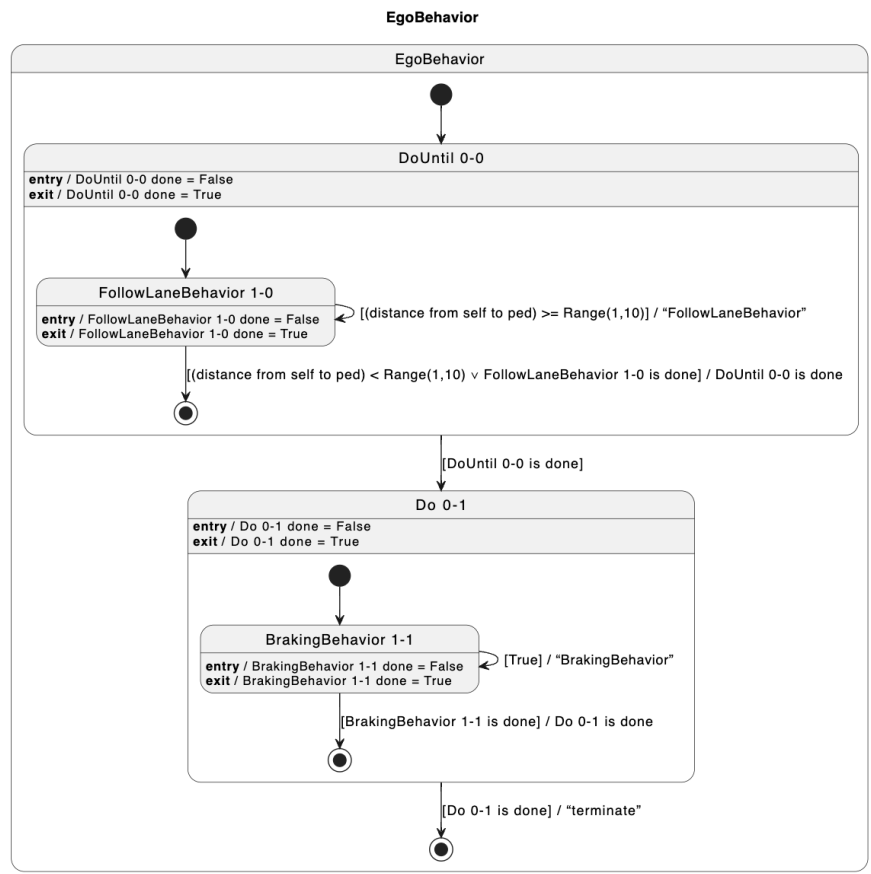

Figure A.4: Statechart representation for a sequence of `do until` and `do` statements. The modular structure of each statement allows them to be composed together, regardless of the statement type. This aims to mirror the SCENIC program execution of one line of code at a time.

# Many Do Statements

**Scenic Code**

```
behavior EgoBehavior():
    do AccelerateForwardBehavior() until (self can see ped)
    do FollowLaneBehavior() until (distance from self to ped) < Range(1,10)
    do Brake() until ped not in self.lane
    do LaneChangeBehavior()

ego = new Car with behavior EgoBehavior()
ped = new Pedestrian
```

**Statechart Representation**

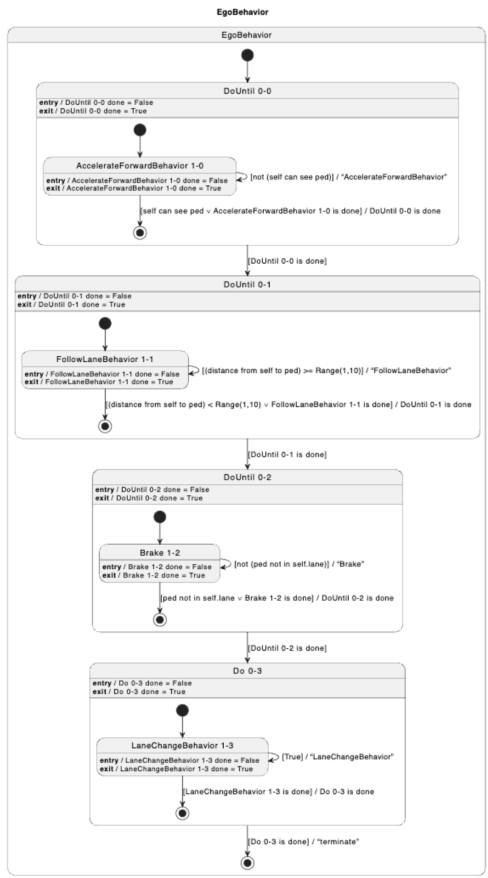

Figure A.5: The modular translation process can scale to any number of statements in sequence.

## Mixing Try and Do Statements

**Scenic Code**

```
behavior EgoBehavior():
    do AccelerateForwardBehavior() until (self can see ped)
    try:
        do FollowLaneBehavior()
    interrupt when (distance from self to ped) < Range(1,10):
        do Brake() until ped not in self.lane
    do LaneChangeBehavior()

ego = new Car with behavior EgoBehavior()
ped = new Pedestrian
```

**Statechart Representation**

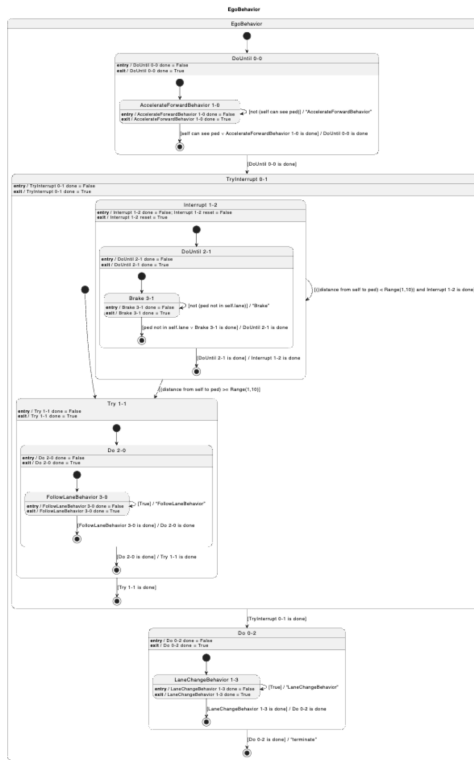

Figure A.6: The `try` statement takes the same format as previously stated in Appendix A.3. We connect several `do` and `try` states in sequence according to the SCENIC program above.

## Nested Sequential Statements

**Scenic Code**

```
behavior EgoBehavior():
    try:
        do AccelerateForwardBehavior() until (self can see ped)
        do FollowLaneBehavior()
    interrupt when (distance from self to ped) < Range(1,10):
        do Brake() until ped not in self.lane
    do LaneChangeBehavior()

ego = new Car with behavior EgoBehavior()
ped = new Pedestrian
```

**Statechart Representation**

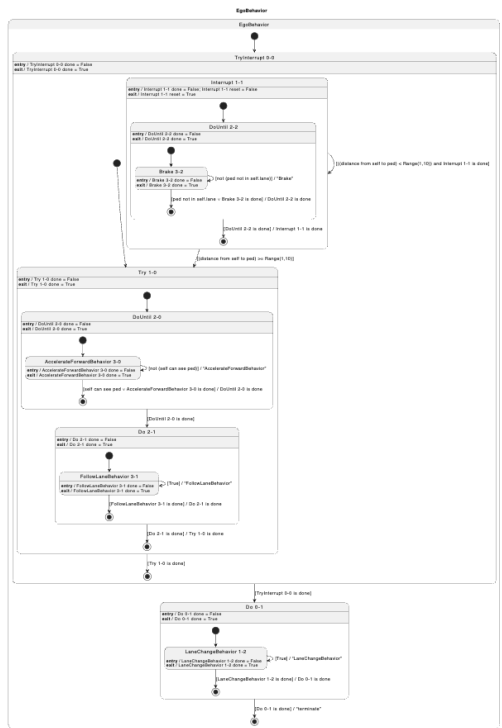

Figure A.7: The `try` statement now contains a sequence of a `do until` and `do` statement in sequence. We now have demonstrated the approaches of handling sequences of SCENIC statements and nested sequences of statements.

## Nested Try Statements

**Scenic Code**

```
behavior EgoBehavior():
    try:
        try:
            do AccelerateForwardBehavior() until (self can see ped)
        interrupt when (distance from self to ped) < Range(1,10):
            do FollowLaneBehavior()
    interrupt when ped in self.lane:
        do BrakingBehavior()

ego = new Car with behavior EgoBehavior()
ped = new Pedestrian
```

**Statechart Representation**

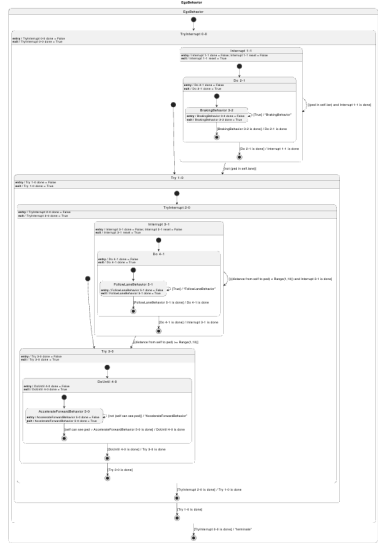

Figure A.8: This nested `try` block demonstrates the capacity of this representation to automatically construct an interrupt-driven hierarchical finite state machine for the specified fragment of SCENIC. The translation process supports unlimited levels of nesting and unlimited numbers of sequential statements without any additional user input. All displayed statecharts representations are generated alongside working UCLID5 modules for BMC.

This marks the conclusion of my thesis: *Querying Labeled Time Series Data with Scenario Programs*. To everyone and everywhere that is Berkeley to me, I hope to see you soon. You will always have a special place in my heart.